\documentclass[aps,prd,groupedaddress,showpacs]{revtex4}
\usepackage{amssymb}
\usepackage{amsmath}
\usepackage{amsfonts}
\usepackage{epsfig}
\usepackage{appendix}
\def\uu{\langle \bar u u \rangle}

\def\ss{\langle \bar s s \rangle}

\newcommand{\seq}{\begin{subequations}}
\newcommand{\sen}{\end{subequations}}
\newcommand{\eq}{\begin{eqnarray}}
\newcommand{\en}{\end{eqnarray}}

\begin{document}

\title{Light cone QCD sum rules for the $g_{\Xi_Q\Xi_Q\pi}$ coupling constant }
\noindent
\author{
       Su Houng Lee$^{1*}$,
       A. Ozpineci$^{2**}$,
        Y. Sarac$^{3***}$,
      }\\

\affiliation{\vspace*{1.2\baselineskip}\\$^1$ Institute of Physics
and Applied Physics, Yonsei University, 120-749, Seoul, Korea\\
$^*$ suhoung@phya.yonsei.ac.kr
\vspace*{1.2\baselineskip} \\$^2$
Physics
Department, Middle East Technical University, 06531, Ankara, Turkey\\
$^**$ ozpineci@metu.edu.tr
\vspace*{1.2\baselineskip} \\
$^3$ Electrical and Electronics Engineering Department, Atilim University, 06836 Ankara, Turkey\\
$^{***}$ ysoymak@atilim.edu.tr \vspace*{1.2\baselineskip} }

\date{\today}

\begin{abstract}
 For the heavy baryons $\Xi_c$ and $\Xi_b$ the coupling constants
 $g_{\Xi_c\Xi_c\pi}$ and $g_{\Xi_b\Xi_b\pi}$ are calculated in the
 framework of light cone QCD sum rules. The most general form of the
 interpolating field of $\Xi_Q$ is used in the calculation.
\end{abstract}


\maketitle

\section{Introduction}
Experimental progresses over the last few years provided exciting
results in the heavy baryon sector. There have been plenty of heavy
baryon observations yielding a large amount of experimental data on
charm and bottom baryons. Observation of Belle, BABAR, DELPHI, CLEO,
CDF, DO etc \cite{Mizuk2005, Mizuk2007, Aubert2007, Feindt2007,
Edwards1995,Artuso2001, I.V.Gorelov2007, Aaltonen2007, Abazov2007}
have motivated the theoretical interests on heavy baryons containing
c and b quarks. The mass spectroscopy of these baryons have been
studied using various theoretical models \cite{Capstick, Roncaglia,
Jenkins, Mathur, Ebert, Karliner1, Karliner2, Rosner, Karliner3} as
well as QCD sum rules method \cite{Shifman, Bagan, Navarra, Shuryak,
Grozin, Dai, Wang1, Zhu1, Huang, Wang2, Wang3, Duraes, Liu}. Their
masses calculated via QCD sum rules in heavy quark limit
\cite{Shuryak} and in the heavy quark effective theory\cite{Grozin,
Dai, Wang1}.

Beside the mass spectrum there have been theoretical studies on the
magnetic moments of these baryons utilizing naive quark model
\cite{Choudhury, Lic}, quark model \cite{Glozman, Julia}, bound
state approach \cite{Schollamd}, relativistic three-quark model
\cite{Faessler}, hyper central model \cite{Patel}, Chiral
perturbation model \cite{Savage}, soliton model \cite{DOR}, skyrmion
model \cite{Oh}, nonrelativistic constituent quark model \cite{An},
QCD sum rules in external magnetic fiels \cite{Zhu2}, and light cone
QCD sum rules (LCQSR)  method \cite{Aliev1, Aliev2}.

In the present work we make use of the (LCQSR) to calculate the
coupling constant $g_{\Xi_Q\Xi_Q\pi}$. A similar work has been done
in \cite{Azizi} for the coupling constant
$g_{\Sigma_Q\Lambda_Q\pi}$. This paper is organized as follows. In
section 2, we introduce the interpolating field for $\Xi_Q$ and give
the details of (LCQSR) calculations for the coupling constant. In
section 3 the numerical analysis, discussion and conclusion are
presented.

\section{ Light cone QCD sum rules for the $g_{\Xi_Q\Xi_Q\pi}$ coupling constant}
To calculate the coupling constant $g_{{\Xi_{Q}\Xi_{Q}\pi}}$  via
the (LCQSR) one studies a suitably chosen correlation function of
the form
\begin{equation}\label{T}
\Pi=i\int d^{4}xe^{ipx}\langle \pi(q)\mid {\cal
T}\{\eta_{\Xi_{Q}}(x)\bar{\eta}_{\Xi_{Q}}(0) \}\mid0\rangle,
\end{equation}
where $\eta_{\Xi_{Q}}(x)$ denotes the interpolating current of
$\Xi_{Q}$ baryon and ${\cal T}$ denotes the time ordering product.
One can calculate this correlation function either
phenomenologically, inserting a complete set of hadronic states into
the correlator to obtain a result containing hadronic parameters, or
theoretically via the operator product expansion (OPE) in deep
Euclidean region $p^2\rightarrow\infty$ in terms of QCD parameters.
Sum rules are obtained by matching these two expressions after Borel
transformations and the contribution of the higher states and
continuum is subtracted.

The calculation of the phenomenological side is similar to the
calculation in \cite{Azizi} and the details are presented for
completeness. To obtain the physical representation of the
correlator a complete set of hadronic state having the quantum
number of $\Xi_{Q}$ baryon is inserted. Then the correlation
function becomes
\begin{eqnarray}\label{T2}
\Pi&=&\frac{\langle0\mid \eta_{\Xi_{Q}}\mid
\Xi_{Q}(p_{2})\rangle}{p_{2}^{2}-m_{\Xi_{Q}}^{2}}\langle
\Xi_{Q}(p_{2}) \pi(q)\mid \Xi_{Q}(p_{1})\rangle\frac{\langle
\Xi_{Q}(p_{1})\mid
\bar{\eta}_{\Xi_{Q}}\mid 0\rangle}{p_{1}^{2}-m_{\Xi_{Q}}^{2}}+...,\nonumber\\
\end{eqnarray}
where $p_1=p+q$ and $p_2=p$ and ... represents the contribution of
the higher states and continuum. The matrix elements representing
the coupling of the interpolating field to the baryon state under
consideration  are defined as
\begin{equation}\label{lambdabey}
\langle0\mid \eta_{\Xi_{Q}}\mid
\Xi_Q(p,s)\rangle=\lambda_{\Xi_Q}u_{\Xi_Q}(p,s),
\end{equation}
where $\lambda_{\Xi_Q}$ denotes the coupling strength and
$u_{\Xi_Q}$ is the spinor for the $\Xi_Q$ baryon. The coupling
constant $g_{\Xi_Q\Xi_Q\pi}$ is defined by the matrix element in
Eq.~(\ref{T2}) which is given as
\begin{eqnarray}\label{matelpar}
\langle \Xi_{Q}(p_{2}){\pi}(q)\mid
\Xi_{Q}(p_{1})\rangle &=&g_{\Xi_{Q}\Xi_{Q}\pi}\overline{u}(p_{2})i\gamma_{5}u(p_{1}).\nonumber\\
\end{eqnarray}
Using the Eqs.~(\ref{lambdabey}) and(\ref{matelpar})in
Eq.~(\ref{T2}) one obtains the phenomenological side of the
correlator as
\begin{eqnarray}\label{final phenpart}
\Pi&=&i\frac{g_{\Xi_{Q}\Xi_{Q}\pi}|\lambda_{\Xi_{Q}}|^2}{(p_{1}^{2}-m_{\Xi_{Q}}^{2})(p_{2}^{2}-m_{\Xi_{Q}}^{2})}
\left[-\not\!p
\not\!q\gamma_{5}-m_{\Xi_{Q}}\not\!q\gamma_{5}\right].
\end{eqnarray}
The coefficient of any one of the structures
$\not\!p\not\!q\gamma_5$ or $\not\!q\gamma_5$ can be used. In this
work, we will work with the structure $\not\!q\gamma_5$.

For the calculation of the QCD side of the correlation function
which is obtained via (OPE) one needs to know the explicit
expression of the interpolating field of $\Xi_Q$ which is given in
the following form:
\begin{eqnarray}\label{currentguy}
\eta_{\Xi_{Q}} & = & \frac{1}{\sqrt{2}}\epsilon^{abc}\Bigg[
\bigg(u_a^T C s_b \bigg) \gamma^5Q_c + \beta \bigg(u_a^T C \gamma^5
s_b \bigg)Q_c +\bigg(u_a^T C Q_b \bigg) \gamma^5s_c + \beta
\bigg(u_a^T C \gamma^5 Q_b \bigg)s_c\Bigg]
\end{eqnarray}
where $Q$ represents the heavy quarks $c$ or $b$, $\beta$ is an
arbitrary parameter with $\beta=-1$ corresponding to the Ioffe
current, $C$ is the charge conjugation operator and $a,b,c$ are the
color indices. After inserting the interpolating fields into
Eq.~(\ref{T}) and carrying out the contractions, the following
expression is obtained:

\begin{eqnarray}\label{contraction}
\Pi& = & -\frac{i}{2}\epsilon^{abc}\epsilon^{a'b'c'}\int d^4x
e^{ipx}\langle\pi(q)\mid\Bigg\{-\gamma_5S_Q^{cb'}CS_u^{Taa'}CS_s^{bc'}\gamma_5-\gamma_5S_s^{cb'}CS_u^{Taa'}CS_Q^{bc'}\gamma_5\nonumber\\&&
+Tr[CS_u^{Taa'}CS_s^{bb'}]\gamma_5
S_Q^{cc'}\gamma_5+Tr[CS_u^{Taa'}CS_Q^{bb'}]\gamma_5
S_s^{cc'}\gamma_5+\beta\Bigg[-\gamma_5S_Q^{cb'}\gamma_5CS_u^{Taa'}CS_s^{bc'}\nonumber\\&&
-\gamma_5S_s^{cb'}\gamma_5CS_u^{Taa'}CS_Q^{bc'}-S_Q^{cb'}CS_u^{Taa'}C\gamma_5S_s^{bc'}\gamma_5
-S_s^{cb'}CS_u^{Taa'}C\gamma_5S_Q^{bc'}\gamma_5\nonumber\\&&+Tr[\gamma_5CS_u^{Taa'}CS_s^{bb'}]\gamma_5
S_Q^{cc'}+Tr[\gamma_5CS_u^{Taa'}CS_Q^{bb'}]\gamma_5
S_s^{cc'}+Tr[CS_u^{Taa'}C\gamma_5S_s^{bb'}]S_Q^{cc'}\gamma_5\nonumber\\&&+Tr[CS_u^{Taa'}C\gamma_5S_Q^{bb'}]
S_s^{cc'}\gamma_5\Bigg]+\beta^2\Bigg[-S_Q^{cb'}\gamma_5CS_u^{Taa'}C\gamma_5S_s^{bc'}
-S_s^{cb'}\gamma_5CS_u^{Taa'}C\gamma_5S_Q^{bc'}\nonumber\\&&+Tr[\gamma_5CS_u^{Taa'}C\gamma_5S_s^{bb'}]S_Q^{cc'}
+Tr[\gamma_5CS_u^{Taa'}C\gamma_5S_Q^{bb'}]S_s^{cc'}\Bigg]\Bigg\}\mid0\rangle.
\end{eqnarray}
Note that this is a schematical representation. The pion can be
emitted from any one of the $u$ or $d$ quarks and hence both
contribution should be summed. To obtain the contribution of pion
emission from any one of the quarks, its propagator is replaced by
$:q^a(0)q^b(x):$. To proceed with the calculation the heavy and
light quark propagators are needed. In this work, the following
propagators are used: \cite{Balitsky}
\begin{eqnarray}\label{heavylightguy}
 S_Q (x)& =&  S_Q^{free} (x) - i g_s \int \frac{d^4 k}{(2\pi)^4}
e^{-ikx} \int_0^1 dv \Bigg[\frac{\not\!k + m_Q}{( m_Q^2-k^2)^2}
G^{\mu\nu}(vx) \sigma_{\mu\nu} + \frac{1}{m_Q^2-k^2} v x_\mu
G^{\mu\nu} \gamma_\nu \Bigg],
\nonumber \\
S_q(x) &=&  S_q^{free} (x)  - \frac{\langle \bar q q \rangle}{12} -
\frac{x^2}{192} m_0^2 \langle \bar q q \rangle  \nonumber \\ &&
 - i g_s \int_0^1 du \left[\frac{\not\!x}{16 \pi^2 x^2} G_{\mu \nu} (ux) \sigma_{\mu \nu} - u x^\mu
G_{\mu \nu} (ux) \gamma^\nu \frac{i}{4 \pi^2 x^2} \right].
 \end{eqnarray}
 where the free light and heavy quark propagators in Eq.~(\ref{heavylightguy}) are given in $x$
 representation as
 \begin{eqnarray}\label{free1guy}
S^{free}_{q} &=&\frac{i\not\!x}{2\pi^{2}x^{4}},\nonumber\\
S^{free}_{Q}
&=&\frac{m_{Q}^{2}}{4\pi^{2}}\frac{K_{1}(m_{Q}\sqrt{-x^2})}{\sqrt{-x^2}}-i
\frac{m_{Q}^{2}\not\!x}{4\pi^{2}x^2}K_{2}(m_{Q}\sqrt{-x^2}),\nonumber\\
\end{eqnarray}
where $K_{i}$ are the Bessel functions. As seen from
Eq.~(\ref{contraction}) as well as the propagators the matrix
elements of the form $\langle\pi(q)|\bar{q}(x_1)\Gamma_i
q(x_2)|0\rangle$ are also needed. Here $\Gamma_i$ represents any
member of the Dirac basis i.e. $\{1, \gamma_{\alpha},
\sigma_{\alpha\beta}/\sqrt{2}, i\gamma_{5}\gamma_{\alpha},
\gamma_{5}\}$. In terms of the pion light cone distribution
amplitudes the matrix elements $\langle\pi(q)|\bar{q}(x_1)\Gamma_i
q(x_2)|0\rangle$ are given explicitly as~\cite{R21,R22} \\*
\begin{eqnarray} \langle {\pi}(p)| \bar
q(x) \gamma_\mu \gamma_5 q(0)| 0 \rangle &=& -i f_{\pi} p_\mu
\int_0^1 du  e^{i \bar u p x}
    \left( \varphi_{\pi}(u) + \frac{1}{16} m_{\pi}^2 x^2 {\mathbb A}(u) \right)
\nonumber \\
    &-& \frac{i}{2} f_{\pi} m_{\pi}^2 \frac{x_\mu}{px} \int_0^1 du e^{i \bar u px} {\mathbb B}(u),
\nonumber \\
\langle {\pi}(p)| \bar q(x) i \gamma_5 q(0)| 0 \rangle &=& \mu_{\pi}
\int_0^1 du e^{i \bar u px} \varphi_P(u),
\nonumber \\
\langle {\pi}(p)| \bar q(x) \sigma_{\alpha \beta} \gamma_5 q(0)| 0
\rangle &=& \frac{i}{6} \mu_{\pi} \left( 1 - \tilde \mu_{\pi}^2
\right) \left( p_\alpha x_\beta - p_\beta x_\alpha\right)
    \int_0^1 du e^{i \bar u px} \varphi_\sigma(u),
\nonumber \\
\langle {\pi}(p)| \bar q(x) \sigma_{\mu \nu} \gamma_5 g_s G_{\alpha
\beta}(v x) q(0)| 0 \rangle &=&
    i \mu_{\pi} \left[
        p_\alpha p_\mu \left( g_{\nu \beta} - \frac{1}{px}(p_\nu x_\beta + p_\beta x_\nu) \right)
\right. \nonumber \\
    &-& p_\alpha p_\nu \left( g_{\mu \beta} - \frac{1}{px}(p_\mu x_\beta + p_\beta x_\mu) \right)
\nonumber \\
    &-& p_\beta p_\mu \left( g_{\nu \alpha} - \frac{1}{px}(p_\nu x_\alpha + p_\alpha x_\nu) \right)
\nonumber \\
    &+& p_\beta p_\nu \left. \left( g_{\mu \alpha} - \frac{1}{px}(p_\mu x_\alpha + p_\alpha x_\mu) \right)
        \right]
\nonumber \\
    &\times& \int {\cal D} \alpha e^{i (\alpha_{\bar q} + v \alpha_g) px} {\cal T}(\alpha_i),
\nonumber \\
\langle {\pi}(p)| \bar q(x) \gamma_\mu \gamma_5 g_s G_{\alpha \beta}
(v x) q(0)| 0 \rangle &=&
    p_\mu (p_\alpha x_\beta - p_\beta x_\alpha) \frac{1}{px} f_{\pi} m_{\pi}^2
        \int {\cal D}\alpha e^{i (\alpha_{\bar q} + v \alpha_g) px} {\cal A}_\parallel (\alpha_i)
\nonumber \\
    &+& \left[
        p_\beta \left( g_{\mu \alpha} - \frac{1}{px}(p_\mu x_\alpha + p_\alpha x_\mu) \right) \right.
\nonumber \\
    &-&     p_\alpha \left. \left(g_{\mu \beta}  - \frac{1}{px}(p_\mu x_\beta + p_\beta x_\mu) \right) \right]
    f_{\pi} m_{\pi}^2
\nonumber \\
    &\times& \int {\cal D}\alpha e^{i (\alpha_{\bar q} + v \alpha _g) p x} {\cal A}_\perp(\alpha_i),
\nonumber
\end{eqnarray}
\begin{eqnarray}
\langle {\pi}(p)| \bar q(x) \gamma_\mu i g_s G_{\alpha \beta} (v x)
q(0)| 0 \rangle &=&
    p_\mu (p_\alpha x_\beta - p_\beta x_\alpha) \frac{1}{px} f_{\pi} m_{\pi}^2
        \int {\cal D}\alpha e^{i (\alpha_{\bar q} + v \alpha_g) px} {\cal V}_\parallel (\alpha_i)
\nonumber \\
    &+& \left[
        p_\beta \left( g_{\mu \alpha} - \frac{1}{px}(p_\mu x_\alpha + p_\alpha x_\mu) \right) \right.
\nonumber \\
    &-&     p_\alpha \left. \left(g_{\mu \beta}  - \frac{1}{px}(p_\mu x_\beta + p_\beta x_\mu) \right) \right]
    f_{\pi} m_{\pi}^2
\nonumber \\
    &\times& \int {\cal D}\alpha e^{i (\alpha_{\bar q} + v \alpha _g) p x} {\cal V}_\perp(\alpha_i),
\end{eqnarray}
where $\mu_{\pi} = f_{\pi} \frac{m_{\pi}^2}{m_{u} + m_{d}},$ $\tilde
\mu_{\pi} = \frac{{m_{u} + m_{d}}}{m_{\pi}}$, ${\cal D} \alpha =
 d \alpha_{\bar q}  d \alpha_q  d \alpha_g
\delta(1-\alpha_{\bar q}-\alpha_q-\alpha_g)$ and the
$\varphi_{\pi}(u),$ $\mathbb A(u),$ $\mathbb B(u),$ $\varphi_P(u),$
$\varphi_\sigma(u),$ ${\cal T}(\alpha_i),$ ${\cal
A}_\perp(\alpha_i),$ ${\cal A}_\parallel(\alpha_i),$ ${\cal
V}_\perp(\alpha_i)$ and ${\cal V}_\parallel(\alpha_i)$ are functions
of definite twist and their expressions will be given in the
numerical analysis section.

With these inputs, the correlation function can be calculated in
terms of quark-gluon degrees of freedom. To match the two
representation, their spectral representation is used. The
contributions of the higher states and continuum are subtracted
using quark-hadron duality. Furthermore, to eliminate the unknown
polynomials in the spectral representation and suppress the
contribution of higher states and continuum, Borel transformation is
applied with respect to $p^2$ and $(p+q)^2$. Finally, the sum rules
is obtained from the integral:

\begin{eqnarray}\label{magneticmoment2}
e^{\frac{-m_{\Xi_Q}}{M^2}}m_{\Xi_Q}|\lambda_{\Xi_Q}|^2g_{\Xi_Q\Xi_Q\pi}&=&
\int_{m_{Q}^{2}}^{s_{0}}e^{\frac{-s}{M^{2}}}\rho(s)ds+e^{\frac{-m_Q^2}{M^{2}}}\Gamma,
\end{eqnarray}
where the explicit expressions of $\rho(s)$ and $\Gamma$ are given
in Appendix A.

To obtain a prediction for $g_{\Xi_Q\Xi_Q\pi}$, the residue
$\lambda_{\Xi_Q}$ is also needed. The residue can be calculated
using mass sum rules and is given as:
\begin{eqnarray}\label{residu2}
-\lambda_{\Xi_{Q}}^{2}e^{-m_{\Xi_{Q}}^{2}/M^{2}}&=&\int_{m_{Q}^{2}}^{s_{0}}e^{\frac{-s}{M^{2}}}\rho_{1}(s)ds
+e^{\frac{-m_Q^2}{M^{2}}}\Gamma_{1},
\end{eqnarray}
where explicit expressions of $\rho_1(s)$ and $\Gamma_1$ are given
in Appendix B. From Eq.~(\ref{residu2}), the mass can be obtained by
differentiation with respect to $M^2$ as
\begin{eqnarray}\label{mass}
m_{\Xi_{Q}}&=&\frac{\int_{m_{Q}^{2}}^{s_{0}}e^{\frac{-s}{M^{2}}}s~\rho_{1}(s)ds
+M^4
\frac{d}{dM^2}(e^{\frac{-m_Q^2}{M^{2}}}\Gamma_{1})}{\int_{m_{Q}^{2}}^{s_{0}}e^{\frac{-s}{M^{2}}}\rho_{1}(s)ds
+e^{\frac{-m_Q^2}{M^{2}}}\Gamma_{1}}.
\end{eqnarray}

\section{Numerical Analysis}
In this section the numerical analysis for the coupling constant
$g_{\Xi_Q\Xi_Q\pi}$ is presented. The required input parameters are
given as: $\uu(1~GeV) = -(0.243)^3~GeV^3$, $\ss(1~GeV) = 0.8
\uu(1~GeV)$, $m_b = 4.7~GeV$, $m_c = 1.23~GeV$, $m_{\Xi_{b}} =
5.792~GeV$, $m_{\Xi_{c}} = 2.470~GeV$, and $m_0^2(1~GeV) =
(0.8\pm0.2)~GeV^2$ \cite{Belyaev}, $f_\pi=0.131$ \cite{R21,Belyaev},
$m_\pi=0.135~GeV$ . We also need the $\pi$-meson wave functions for
the coupling constant calculation, whose explicit forms are
presented as \cite{R21, R22}
\begin{eqnarray}
\phi_{\pi}(u) &=& 6 u \bar u \left( 1 + a_1^{\pi} C_1(2 u -1) +
a_2^{\pi} C_2^{3 \over 2}(2 u - 1) \right),
\nonumber \\
{\cal T}(\alpha_i) &=& 360 \eta_3 \alpha_{\bar q} \alpha_q
\alpha_g^2 \left( 1 + w_3 \frac12 (7 \alpha_g-3) \right),
\nonumber \\
\phi_P(u) &=& 1 + \left( 30 \eta_3 - \frac{5}{2}
\frac{1}{\mu_{\pi}^2}\right) C_2^{1 \over 2}(2 u - 1)
\nonumber \\
&+& \left( -3 \eta_3 w_3  - \frac{27}{20} \frac{1}{\mu_{\pi}^2} -
\frac{81}{10} \frac{1}{\mu_{\pi}^2} a_2^{\pi} \right)
C_4^{1\over2}(2u-1),
\nonumber \\
\phi_\sigma(u) &=& 6 u \bar u \left[ 1 + \left(5 \eta_3 - \frac12
\eta_3 w_3 - \frac{7}{20}  \mu_{\pi}^2 - \frac{3}{5} \mu_{\pi}^2
a_2^{\pi} \right) C_2^{3\over2}(2u-1) \right],
\nonumber \\
{\cal V}_\parallel(\alpha_i) &=& 120 \alpha_q \alpha_{\bar q}
\alpha_g \left( v_{00} + v_{10} (3 \alpha_g -1) \right),
\nonumber \\
{\cal A}_\parallel(\alpha_i) &=& 120 \alpha_q \alpha_{\bar q}
\alpha_g \left( 0 + a_{10} (\alpha_q - \alpha_{\bar q}) \right),
\nonumber
\end{eqnarray}
\begin{eqnarray}
{\cal V}_\perp (\alpha_i) &=& - 30 \alpha_g^2\left[
h_{00}(1-\alpha_g) + h_{01} (\alpha_g(1-\alpha_g)- 6 \alpha_q
\alpha_{\bar q}) +
    h_{10}(\alpha_g(1-\alpha_g) - \frac32 (\alpha_{\bar q}^2+ \alpha_q^2)) \right],
\nonumber\\
{\cal A}_\perp (\alpha_i) &=& 30 \alpha_g^2(\alpha_{\bar q} -
\alpha_q) \left[ h_{00} + h_{01} \alpha_g + \frac12 h_{10}(5
\alpha_g-3) \right],
\nonumber \\
B(u)&=& g_{\pi}(u) - \phi_{\pi}(u),
\nonumber \\
g_{\pi}(u) &=& g_0 C_0^{\frac12}(2 u - 1) + g_2 C_2^{\frac12}(2 u -
1) + g_4 C_4^{\frac12}(2 u - 1),
\nonumber \\
{\mathbb A}(u) &=& 6 u \bar u \left[\frac{16}{15} + \frac{24}{35}
a_2^{\pi}+ 20 \eta_3 + \frac{20}{9} \eta_4 +
    \left( - \frac{1}{15}+ \frac{1}{16}- \frac{7}{27}\eta_3 w_3 - \frac{10}{27} \eta_4 \right) C_2^{3 \over 2}(2 u - 1)
    \right. \nonumber \\
    &+& \left. \left( - \frac{11}{210}a_2^{\pi} - \frac{4}{135} \eta_3w_3 \right)C_4^{3 \over 2}(2 u - 1)\right]
\nonumber \\
&+& \left( -\frac{18}{5} a_2^{\pi} + 21 \eta_4 w_4 \right)\left[ 2
u^3 (10 - 15 u + 6 u^2) \ln u
\right. \nonumber \\
&+& \left. 2 \bar u^3 (10 - 15 \bar u + 6 \bar u ^2) \ln\bar u + u
\bar u (2 + 13 u \bar u) \right] \label{wavefns},
\end{eqnarray}
where $C_n^k(x)$ are the Gegenbauer polynomials,
\begin{eqnarray}
h_{00}&=& v_{00} = - \frac13\eta_4,
\nonumber \\
a_{10} &=& \frac{21}{8} \eta_4 w_4 - \frac{9}{20} a_2^{\pi},
\nonumber \\
v_{10} &=& \frac{21}{8} \eta_4 w_4,
\nonumber \\
h_{01} &=& \frac74  \eta_4 w_4  - \frac{3}{20} a_2^{\pi},
\nonumber \\
h_{10} &=& \frac74 \eta_4 w_4 + \frac{3}{20} a_2^{\pi},
\nonumber \\
g_0 &=& 1,
\nonumber \\
g_2 &=& 1 + \frac{18}{7} a_2^{\pi} + 60 \eta_3  + \frac{20}{3}
\eta_4,
\nonumber \\
g_4 &=&  - \frac{9}{28} a_2^{\pi} - 6 \eta_3 w_3 \label{param0}.
\end{eqnarray}

The constants in the Eqs.~(\ref{wavefns}) and (\ref{param0}) are
calculated at the renormalization scale $\mu=1 ~~GeV^{2}$ and are
given as $a_{1}^{\pi} = 0$, $a_{2}^{\pi} = 0.44$, $\eta_{3} =0.015$,
$\eta_{4}=10$, $w_{3} = -3$ and $ w_{4}= 0.2$.

Looking at the result of LCQCD sum rules calculation for the
coupling constant $g_{\Xi_Q\Xi_Q\pi}$ one encounters three auxiliary
parameters. These parameters are the Borel mass $M^2$, the continuum
threshold $s_0$ and the arbitrary parameter $\beta$ and there should
be no dependency of a physical quantity, such as the coupling
constant for our case, on them. Therefore at this stage a working
region of these auxiliary parameters should be determined. In order
to determine the upper and lower bound of $M$ we use the
requirements that the continuum contribution be less than that of
the ground state, and the highest power of $1/M^2$ be less than
$30^0/_{0}$  of the highest power of $M$. The former (latter) is
used to determine upper (lower) bound of $M^2$. To determine the
value of continuum threshold $s_0$ we use the two-point correlation
function from which we obtain the mass sum rules. In Fig.~1 and
Fig.~2, we plot the dependence of our prediction on $m_{\Xi_c}$ to
the Borel parameter $M^2$, and $\cos\theta$, where
$\beta=\cos\theta$, respectively. For these plots, the continuum
threshold is chosen to be $s_0=8$~Gev$^2$ and $s_0=9$~Gev$^2$. For
these values of the continuum threshold we see from these figures
that our predictions are in agreement with experimental results and
that there is no dependence on the auxiliary parameters. In Figs.~3,
4, we carry out the same analysis for $m_{\Xi_b}$ and find the
continuum threshold to be $s_0=36$~Gev$^2$ and $s_0=37$~Gev$^2$. In
Fig.~4, we also observe that our prediction is stable with respect
to variations of $\theta$ for the region $0.5<\cos\theta<0.5$ which
corresponds to $|\beta|>1.7$.

The results for the coupling constant calculation  are presented in
Figs.~5, 6, 7 and 8. Figs. 5 and 7 depicts respectively the
dependence of the coupling constants $g_{\Xi_c\Xi_c\pi}$ and
$g_{\Xi_b\Xi_b\pi}$ on $M^2$ in the working region of $M^2$. The
results are given for two fixed values of $\beta$ and two fixed
values of $s_0$ for each coupling constant. It follows from the
figures that the results are rather stable with respect to the
variations of $M^2$ in the given region of $M^2$. The dependence of
the coupling constants on $\cos\theta$ are also presented in Figs.~6
and 8 and with respect to these figures when $\cos\theta$ is in
between $-0.25<\cos\theta<0.25$ the coupling constant
$g_{\Xi_c\Xi_c\pi}$ is practically independent of the unphysical
parameter $\beta$. The interval of $\beta$ that gives coupling
constant result independent of $\beta$ for $g_{\Xi_b\Xi_b\pi}$ is
$-0.75<\cos\theta<0.75$. As a result of our analysis we obtain the
values of the coupling constants as
\begin{eqnarray}
g_{\Xi_c\Xi_c\pi}=1.0\pm0.5 ,~~~~~~~~~~~~~~~
g_{\Xi_b\Xi_b\pi}=1.6\pm0.4\nonumber
 \label{couplings}.
\end{eqnarray}

To summarize, in this work we present the results of the coupling
constant for the coupling constants $g_{\Xi_c\Xi_c\pi}$ and
$g_{\Xi_b\Xi_b\pi}$. To this end, we make use of the LCQSR approach
with the $\Xi_Q$ current applied in its most general form. We obtain
the appropriate values of the threshold parameters $s_0$ from the
mass sum rules and, using them and appropriate intervals of Borel
parameter and $\beta$, we attain the coupling constants
$g_{\Xi_c\Xi_c\pi}$ and $g_{\Xi_b\Xi_b\pi}$.

\section{Acknowledgment}
The authors  would like to thank T. M. Aliev for his useful
discussions. S.~H.~L. was supported by the Korean Research
Foundation KRF-2006-C00011. The work of A. O. and Y. S. have been
supported in part by the European Union (HadronPhysics2 project
"Study of strongly interacting matter").

\newpage

\section*{Appendix A}
\begin{eqnarray}\label{rho1}
\rho(s)&=&-\frac{f_\pi
m_Q^2}{768\sqrt{2}\pi^2}\Bigg[3m_b\Bigg\{m_Q(1+6\beta+\beta^2)(2\psi_{10}-\psi_{20}
+\psi_{30})-4m_Q(3-4\beta+\beta^2)\psi_{10}\nonumber\\&&-2[m_Q(1+6\beta+\beta^2)\psi_{00}-2m_s(3-4\beta+\beta^2)]
ln(\frac{s}{m_Q^2})\Bigg\}+u_0^2m_{\pi}^2
8(1+7\beta+\beta^2)\psi_{32}\Bigg]\varphi_{\pi}(u_{0})\nonumber\\&&-\frac{
\mu_\pi}{384\sqrt{2}\pi^2}(\beta-1)(\widetilde{\mu}^2-1)\Bigg[2m_Q^2[(-m_Q+3\beta
m_Q+5m_s+3\beta m_s)\psi_{10}
+m_Q(1-3\beta)ln(\frac{s}{m_Q^2})]\nonumber\\&&-[m_Q(3\beta-1)(2\psi_{10}-\psi_{11}-\psi_{12}+2\psi_{21})+m_s(5+3\beta)(\psi_{11}+\psi_{12})]u_0
m_\pi^2\Bigg]\varphi_{\sigma}(u_{0})\nonumber\\&&-\frac{f_\pi
m_\pi^2}{256\sqrt{2}\pi^2}\Bigg[-m_Q\Bigg\{m_Q(1+6\beta+\beta^2)\psi_{10}+m_s(3-4\beta+\beta^2)\psi_{00}\Bigg\}\nonumber\\&&
+2(1+7\beta+\beta^2)(\psi_{11}+\psi_{12})u_0^2
m_\pi^2\Bigg]A(u_0)-\frac{f_\pi m_Q^4
u_0}{384\sqrt{2}\pi^2}[1+7\beta+\beta^2)\Bigg[6\psi_{10}-3\psi_{20}+\psi_{30}-2\psi_{41}\nonumber\\&&
-6\psi_{00}ln(\frac{s}{m_Q^2})\Bigg]\varphi_{\pi}'(u_{0})-\frac{\mu_\pi
m_Q^2u_0}{768\sqrt{2}\pi^2}(\beta-1)(\widetilde{\mu}^2-1)\Bigg[m_Q(3\beta-1)(2\psi_{10}-\psi_{20}+\psi_{31})
\nonumber\\&&+m_s(5+3\beta)(\psi_{20}-\psi_{31})+2m_Q(1-3\beta)ln(\frac{s}{m_Q^2})\Bigg]\varphi_{\sigma}'(u_{0})-\frac{f_\pi
m_\pi^2 m_Q^2
u_0}{256\sqrt{2}\pi^2}(1+7\beta+\beta^2)(\psi_{20}-\psi_{31})A'(u_0)\nonumber\\&&-\frac{f_\pi
\mu_\pi}{64\sqrt{2}\pi^2m_Q}\Bigg[m_s(-1-4\beta+5\beta^2)[m_Q^2(-1+3\psi_{00}-\psi_{01}+\psi_{10})-s\psi_{00}]
(\eta_1-2\eta_2)ln(\frac{\Lambda^2}{m_Q})\nonumber\\&&
+m_Q^2\mu_\pi\Bigg\{m_s(-1-4\beta+5\beta^2)\psi_{01}(\eta_1-2\eta_2)ln(\frac{s}{\Lambda^2})
+[m_Qu_0(2+14\beta+2\beta^2)(2\eta'_4-\eta'_3)\nonumber\\&&-(5+14\beta+5\beta^2)(2\eta'_8-\eta_7)
+m_Q(7-2\beta+7\beta^2)\eta_1+m_s(-5+4\beta+\beta^2)(\eta_1-2\eta_2)]ln(\frac{s}{m_Q^2})\nonumber\\&&
+m_s(-1-4\beta+5\beta^2)(\eta_1-2\eta_2)\Bigg(\gamma_E(2\psi_{00}-\psi_{10}+\psi_{11}-\psi_{21})+(-2\psi_{00}
+\psi_{01}+\psi_{10}-\psi_{21})ln(\frac{s-m_Q^2}{\Lambda^2})\nonumber\\&&
-\psi_{00}ln(\frac{m_Q^2(s-m_Q^2)}{\Lambda^2s})\Bigg)\Bigg\}-[m_Q^4(1-4\beta+3\beta^2)
+m_Q^3m_s(-5+2\beta+3\beta^2)](2\eta'_6-\eta'_5)ln(\frac{s}{m_Q^2})\Bigg]\nonumber\\&&+\frac{1}{128\sqrt{2}\pi^2}\Bigg[f_\pi
m_\pi^2\mu_\pi\Bigg\{2m_Q\Bigg(3m_Qu_0(1+\beta^2)(\psi_{20}-\psi_{31})(2\eta'_8-\eta'_7)
+m_Q(1+\beta^2)[(2\eta'_8-\eta_7'-10\eta'_4+5\eta'_3)\nonumber\\&&+14(1+\beta^2)m_Q\eta_1
-m_s(1-5\beta^2)(\eta_1-2\eta_2)]\psi_{10}+m_Q(1-5\beta^2)(\psi_{00}
-\psi_{01}-\psi_{21})(\eta_1-2\eta_2)\Bigg)\nonumber\\&&+3m_Qu_0[m_Q(\psi_{20}-\psi_{31})-2m_s(\psi_{10}+\psi_{21})]\zeta
+m_Qu_0\beta^2[3m_Q(\psi_{20}-\psi_{31})-2m_s(\psi_{10}+\psi_{21})]\zeta\nonumber\\&&
-2u_0^2m_\pi^2(1+\beta^2)[(12\psi_{10}-3\psi_{11}-3\psi_{12}+12\psi_{21})\eta_1
+(-10\psi_{10}+6\psi_{11}+6\psi_{12}-10\psi_{21})\eta_2]\Bigg\}\nonumber\\&&
+2\mu_\pi[(m_Q-5m_s)+3\beta^2(m_Q+m_s)][(2\eta'_6-\eta'_5)\psi_{10}+2u_0m_\pi^2(\psi_{10}
+\psi_{21})(2\eta_6-\eta_5)]\Bigg]\nonumber\\&&+\frac{\beta
}{64\sqrt{2}\pi^2}\Bigg[f_\pi m_\pi^2
2m_Q\Bigg\{[7m_qu_0(2\eta'_4-\eta'_3+\eta'_7-2\eta'_8)
+4m_Q\eta_1\nonumber\\&&-2m_s(\eta_1-2\eta_2)]\psi_{10}+2m_s(\psi_{00}-\psi_{01}-\psi_{21})(\eta_1-2\eta_2)
+m_Qu_0[11m_Q(\psi_{20}-\psi_{31})+4m_s(\psi_{10}+\psi_{21})]\zeta\nonumber\\&&-2u_0^2m_\pi^2[(6\psi_{10}+3\psi_{11}
+3\psi_{12}+6\psi_{21})\eta_1-14(\psi_{10}+\psi_{21})\eta_2]\Bigg\}+2\mu_\pi\Bigg\{(2m_q-m_s)[m_Q^2(2\eta'_6
-\eta'_5)\nonumber\\&&+2u_0m_\pi^2(\psi_{10}+\psi_{21})(2\eta_6-\eta_5)]\Bigg\}\Bigg]+\frac{
\langle\bar{s}s\rangle}{288\sqrt{2}}\Bigg[3f_\pi\Bigg\{2m_Q(3-4\beta+\beta^2)
-m_s(1+6\beta+\beta^2)\Bigg\}\varphi_{\pi}(u_{0})\nonumber\\&&
+2\mu_\pi\Bigg\{(-5+2\beta+3\beta^2)(\widetilde{\mu}_\pi^2-1)\psi_{00}\Bigg\}\varphi_{\sigma}(u_{0})+6f_\pi
u_0\Bigg\{(1+7\beta+\beta^2)\psi_{00}\Bigg\}\varphi_{\pi}'(u_{0})\nonumber\
\end{eqnarray}
\begin{eqnarray}
&&+\mu_\pi
u_0\Bigg\{(-5+2\beta+3\beta^2)(\widetilde{\mu}_\pi^2-1)\psi_{00}\Bigg\}\varphi_{\sigma}'(u_{0})
-6(1+\gamma_E)\mu_\pi(-5+2\beta+3\beta^2)(2\eta'_6-\eta'_5)\psi_{00}\Bigg]\nonumber\\&&+\frac{
\langle
g_s^2G^2\rangle}{4608\sqrt{2}\pi^2}\Bigg[f_\pi\Bigg\{\frac{-6m_Q^3
m_s}{M^2}(3-4\beta+\beta^2)ln(\frac{s-m_Q^2}{\Lambda^2})\psi_{00}
+\frac{1}{m_Q^2}[-m_Q(1+6\beta+\beta^2)\nonumber\\&&+2m_s(1-9\gamma_E)(3-4\beta+\beta^2)]\psi_{00}
+18m_Qm_s(3-4\beta+\beta^2)(\psi_{00}-\psi_{02}-2\psi_{10}+2\psi_{21}\psi_{22})ln(\frac{s-m_Q^2}{\Lambda^2})
\nonumber\\&&+4u_0^2m_\pi^2(1+7\beta+\beta^2)\psi_{02}\Bigg\}\varphi_{\pi}(u_{0})+\Bigg\{\frac{\mu_\pi
m_s(\tilde{\mu}_\pi^2-1)}{M^2}(-5+2\beta+3\beta^2)ln(\frac{s-m_Q^2}{\Lambda^2})\psi_{10}[u_0^2m_\pi^2(-\frac{3}{M^2}
-\frac{2}{m_Q^2}\nonumber\\&&-\frac{2m_Q^2}{M^4})+\frac{2m_Q^2}{M^4}]
-\frac{\mu_\pi(\beta-1)(\tilde{\mu}_\pi^2-1)}{6m_Q^6}\Bigg(m_Q^4\Big[4m_Q(3\beta-1)\psi_{00}
-3m_s(5+3\beta)(56\gamma_E\psi_{00}-\psi_{20}+\psi_{21}+\psi_{31})\nonumber\\&&+12m_s(5+3\beta)(14\psi_{00}
-14\psi_{03}-42\psi_{10}+30\psi_{20}+27\psi_{21}+18\psi_{22}+9\psi_{23}+24\psi_{31}
+12\psi_{32})ln(\frac{s-m_Q^2}{\Lambda^2})\Big]\nonumber\\&&-6u_0^2m_\pi^2\Big[m_Q^2[m_Q(3\beta-1)\psi_{02}
+m_s(5+3\beta)(-61\psi_{00}+20\psi_{01}+20\psi_{02}+20\psi_{03}+122\psi_{10}+2\psi_{12}+5\psi_{13}
+11\psi_{14}\nonumber\\&&-122\psi_{21}-61\psi_{22}-20\psi_{23})]
+6m_s(5+3\beta)\{[m_Q^2(39\psi_{00}+40\psi_{10})-40s\psi_{00}]ln(\frac{s-m_Q^2}{\Lambda^2})
\nonumber\\&&+m_Q^2\psi_{04}ln(\frac{s(s-m_Q^2)}{m_Q^2\Lambda^2})\}\Big]\Bigg)\Bigg\}\varphi_{\sigma}(u_{0})
+(3-4\beta+\beta^2)f_\pi m_\pi^2
m_s\Bigg\{(\frac{3m_Q^5}{2M^8}-\frac{3m_Q^3}{4M^6})ln(\frac{s-m_Q^2}{\Lambda^2})\nonumber\\&&
+18\frac{1}{m_Q^3}[2\psi_{00}-\psi_{01}-\psi_{02}-3\psi_{10}+3\psi_{21}+\psi_{22}
+2(\psi_{00}-\psi_{03}-3\psi_{10}+3\psi_{21}+2\psi_{22}
+\psi_{23})ln(\frac{s-m_Q^2}{\Lambda^2})]\Bigg\}A(u_0)\nonumber\\&&
-2(1+7\beta+\beta^2)f_\pi
u_0(\psi_{10}+\psi_{21})\varphi_{\pi}'(u_{0})+\Bigg\{\frac{(-5+2\beta+3\beta^2)}{M^2}m_su_0\mu_\pi(\tilde{\mu}_\pi^2-1)(\frac{m_Q^2}{M^2}
+\frac{3}{2})ln(\frac{s-m_Q^2}{\Lambda^2})\psi_{00}\nonumber\\&&
+\frac{1}{6m_Q^2}(\beta-1)(\tilde{\mu}_\pi^2-1)\Bigg((3\beta-1)m_Q(\psi_{00}-3\psi_{10}-3\psi_{21})
+3m_s(5+3\beta)[(20\gamma_E-1)\psi_{00}+\psi_{10}-2\psi_{12}-\psi_{20}+\psi_{31}]\nonumber\\&&
+12m_s(5+3\beta)(5\psi_{00}-5\psi_{03}-15\psi_{10}+30\psi_{20}+24\psi_{31}
+12\psi_{32}+4\psi_{33})ln(\frac{s-m_Q^2}{\Lambda^2})\Bigg)\Bigg\}\varphi_{\sigma}'(u_{0})\nonumber\\&&
+\frac{1}{M^4}(\beta-1)m_Q m_s\Bigg\{3f_\pi
m_\pi^2(7\beta-13)(\eta_1-2\eta_2)-m_Q
\mu_\pi(10+6\beta)(2\eta'_6-\eta'_5)-\frac{2m_Q}{M^2}\mu_\pi^2[m_Qf_\pi(2\beta-14)(\eta_1-2\eta_2)\nonumber\\&&
-u_0(3\beta-9)\zeta]+6u_0(5+3\beta)(2\eta_6-\eta_5)\Bigg\}ln(\frac{s-m_Q^2}{\Lambda^2})-\frac{2}{m_Q^4}(\beta-1)\Bigg\{3m_Qm_sf_\pi\mu_\pi^2
\Bigg((18-6\beta)u_0\zeta
[2\psi_{00}-\psi_{01}\nonumber\\&&-\psi_{02}-3\psi_{10}+3\psi_{21}+\psi_{22}+2(\psi_{00}-\psi_{03}-3\psi_{10}+3\psi_{21}+2\psi_{22}
+\psi_{23})ln(\frac{s-m_Q^2}{\Lambda^2})]+(\eta_1-2\eta_2)\Big[(1+5\beta)\gamma_E\psi_{00}\nonumber\\&&+(15+5\beta)(\psi_{10}-\psi_{11}-\psi_{21})
-(\psi_{00}+3\psi_{02}-4\psi_{03}-36\psi_{10}+15\psi_{11}+15\psi_{12}+36\psi_{21}+20\psi_{22}+4\psi_{23}\nonumber\\&&
+\beta[5\psi_{00}+15\psi_{02}-20\psi_{03}-36\psi_{10}+3\psi_{11}+3\psi_{12}+36\psi_{21}+28\psi_{22}+20\psi_{23}])ln(\frac{s-m_Q^2}{\Lambda^2})\Big]\Bigg)
\nonumber\\&&+\mu_\pi
\Bigg(m_Q^2(\eta'_5-2\eta'_6)\Big[m_Q(3\beta-1)\psi_{00}-18m_s(5+3\beta)[\gamma_E\psi_{00}-(\psi_{00}
-\psi_{02}-2\psi_{10}+3\psi_{20}+3\psi_{31}+\psi_{32})ln(\frac{s-m_Q^2}{\Lambda^2})]\Big]\nonumber\\&&-36m_\pi^2
m_su_0(5+3\beta)(\eta_5-2\eta_6)\Big[2\psi_{00}-\psi_{01}-\psi_{02}-3\psi_{10}+3\psi_{21}+\psi_{22}
+2(\psi_{00}-\psi_{03}-3\psi_{10}+3\psi_{21}+2\psi_{22}\nonumber\\&&+\psi_{23})ln(\frac{s-m_Q^2}{\Lambda^2})\Big]\Bigg)\Bigg\}\Bigg]
,
\end{eqnarray}
and
\begin{eqnarray}\label{gamma1}
\Gamma&=&\frac{f_\pi m_\pi^2}{32\sqrt{2}\pi^2}m_Q m_s\gamma_E
M^2(-1+4\beta+5\beta^2)(\eta_1-2\eta_2)+\frac{\langle\bar{s}s\rangle}{1728\sqrt{2}}\Bigg[f_\pi\Bigg\{\frac{3}{M^4}m_0^2m_Q^2m_s[m_Q^2(1+6\beta+\beta^2)
\nonumber\\&&-4u_0^2m_\pi^2(1+7\beta+\beta^2)]+6m_Q[m_0^2(\beta^2-1)+3m_Qm_s(1+6\beta+\beta^2)]-72m_s
u_0^2m_\pi^2(1+7\beta+\beta^2)\nonumber\\&&+\frac{m_0^2}{M^2}[9m_Q^3(3-4\beta+\beta^2)+m_Q^2m_s(-11+10\beta-11\beta^2)
+2m_su_0^2m_\pi^2(13+46\beta+13\beta^2)]\Bigg\}\varphi_{\pi}(u_{0})\nonumber\\&&
+\mu_\pi(\tilde{\mu}_\pi^2-1)\Bigg\{12(-5+2\beta+3\beta^2)(m_Q^2-u_0^2m_\pi^2)
+\frac{1}{M^2}(\beta-1)\Bigg(\frac{m_0^2m_Q^3m_s}{M^4}(3\beta-1)(m_Q^2-u_0^2m_\pi)\nonumber\\&&
-\frac{m_0^2m_Q}{M^2}[m_Q^4(15+9\beta)+m_Q^2m_s(7\beta-5)-m_Qu_0^2m_\pi^2(15+3\beta)-m_su_0^2m_\pi^2(\beta-3)]-6m_Q^3m_s(3\beta-1)\nonumber\\&&
-m_Q^2m_0^2(\beta-1)+m_Qm_0^2m_s(\beta-3)+2u_0^2m_\pi^2[m_0^2(5+7\beta)+m_Qm_s(9\beta-3)]\Bigg)\Bigg\}\varphi_{\sigma}(u_{0})\nonumber\\&&
+\frac{m_\pi^2f_\pi}{M^2}\Bigg\{\frac{3m_0^2m_Q^4m_s}{4M^6}[m_Q^2(1+6\beta+\beta^2)-4u_0^2m_\pi^2(1+7\beta+\beta^2)]
+\frac{3}{2}[m_Qm_0^2(\beta^2-1)+6m_Q^3(3-4\beta+\beta^2)\nonumber\\&&+12m_su_0^2m_\pi^2(1+7\beta+\beta^2)]
+\frac{m_Q^2}{2M^2}[-9m_Q^2m_s(1+6\beta+\beta^2)+6m_0^2m_Q(5-6\beta+\beta^2)m_0^2m_s(-11+10\beta-11\beta^2)\nonumber\\&&
+36m_su_0^2m_\pi^2(1+7\beta+\beta^2)]+\frac{m_Q^2m_0^2}{4M^4}[-9m_Q^3(3-4\beta+\beta^2)
+2m_Q^2m_s(1-32\beta+\beta^2)\nonumber\\&&-2m_su_0^2m_\pi^2(7+4\beta+7\beta^2)]\Bigg\}A(u_0)
-m_0^2m_su_0f_\pi\Bigg\{(13+46\beta+13\beta^2)+6\frac{m_Q^2}{M^2}(1+7\beta+\beta^2)\Bigg\}\varphi_{\pi}'(u_{0})\nonumber\\&&
+\mu_\pi(\tilde{\mu}_\pi^2-1)u_0(\beta-1)\Bigg\{-m_0^2(7+5\beta)-3m_Qm_s(3\beta-1)+\frac{m_0^2m_Q^3m_s}{2M^4}(3\beta-1)
\nonumber\\&&-\frac{m_0^2m_Q}{2M^2}[3m_Q(5+3\beta)+m_s(\beta-3)]\Bigg\}\varphi_{\sigma}'(u_{0})+m_su_0f_\pi
m_\pi^2\Bigg\{-9(1+7\beta+\beta^2)(1+\frac{m_Q^2}{M^2}-\frac{m_0^2m_Q^4}{6M^6})\nonumber\\&&
+\frac{4m_0^2m_Q^2}{M^4}(7+4\beta+7\beta^2)\Bigg\}A'(u_0)-\frac{3m_0^2m_Q^2m_sf_\pi
m_\pi^2}{M^6}\Bigg\{m_Q^2[(7-2\beta+7\beta^2)\eta_1+u_0(3+22\beta+3\beta^2)\zeta]\nonumber\\&&-u_0^2m_\pi^2[6(1+\beta+\beta^2)\eta_1
-(5+14\beta+5\beta^2)\eta_2]\Bigg\}+\frac{3}{M^2}\Bigg\{f_\pi
m_\pi^2\Bigg(m_Qm_0^2[3(-5+4\beta+\beta^2)(\eta_1-2\eta_2)\nonumber\\&&-2u_0(\beta^2-1)\zeta]
+6m_Q^2m_s[(7-2\beta+7\beta^2)\eta_1+u_0(3+22\beta+3\beta^2)\zeta]-12m_Qm_su_0^2m_\pi^2[6(1+\beta+\beta^2)\eta_1
\nonumber\\&&-(5+14\beta+5\beta^2)\eta_2]\Bigg)+3m_0^2m_Q^2\mu_\pi(-5+2\beta+3\beta^2)(2\eta'_6-\eta'_5)\Bigg\}\nonumber\\&&
-\frac{3m_0^2m_Q^2m_\pi^2}{M^4}\Bigg\{f_\pi\Bigg(3m_Q(-5+4\beta+\beta^2)(\eta_1-2\eta_2)+3m_Qu_0(3-4\beta+\beta^2)\zeta
+2m_su_0(1+7\beta+\beta^2)(2\eta'_4-\eta'_3)\nonumber\\&&-m_s(5+14\beta+5\beta^2)\eta'_7
-2m_s(7-2\beta+7\beta^2)\eta_1+m_su_0(1+6\beta+\beta^2)\zeta\Bigg)-6u_0\mu_\pi(-5+2\beta+3\beta^2)(2\eta_6-\eta_5)\Bigg\}\nonumber\\&&
+18m_\pi^2\Bigg\{f_\pi\Bigg(2m_su_0(1+7\beta+\beta^2)(2\eta'_4-\eta'_3)-m_su_0(5+14\beta+5\beta^2)\eta'_7
+m_su_0(3+22\beta+3\beta^2)\zeta\nonumber\\&&+2m_Q(-5+4\beta+\beta^2)(\eta_1-2\eta_2)+2m_Qu_0(3-4\beta+\beta^2)\zeta\Bigg)
-4u_0\mu_\pi(-5+2\beta+3\beta^2)(2\eta_6-\eta_5)\Bigg\}
\Bigg]\nonumber\\&&+\frac{\langle
g^2G^2\rangle}{4608\sqrt{2}\pi^2}\Bigg[m_sf_\pi
(3-4\beta+\beta^2)\Bigg\{m_Q+\frac{18\gamma_EM^2}{m_Q}-\frac{m_Q^3}{M^2}[2
+3ln(\frac{\Lambda^2}{m_Q^2})]\Bigg\}\varphi_{\pi}(u_{0})\nonumber\\&&
+\mu_\pi
(\tilde{\mu}\pi^2-1)\Bigg\{m_s(-5+2\beta+3\beta^2)\Bigg(-\frac{4M^2\gamma_E}{m_Q^2}
+\frac{m_Q^2}{3M^4}[2+3ln(\frac{\Lambda^2}{m_Q^2})](m_Q^2-u_0^2m_\pi^2)
+\frac{1}{3M^2}[10m_Q^2-6u_0^2m_\pi^2\nonumber\\&&-3u_0^2m_\pi^2\gamma_E-6u_0^2m_\pi^2ln(\frac{\Lambda^2}{m_Q^2})]\Bigg)
+\frac{(\beta-1)}{3m_Q^2}\Bigg(m_Q^2[m_Q(3\beta-1)+9m_s(5+3\beta)]-u_0^2m_\pi^2[m_Q(3\beta-1)\nonumber\
\end{eqnarray}
\begin{eqnarray}
&&+6m_s(5+3\beta)(1+2\gamma_E)+6m_s(5+3\beta)ln(\frac{\Lambda^2}{m_Q^2})]\Bigg)\Bigg\}\varphi_{\sigma}(u_{0})
+f_\pi
m_\pi^2\Bigg\{m_s(3-4\beta+\beta^2)\Bigg(-\frac{3}{4m_Q}\nonumber\\&&
+\frac{m_Q^5}{4M^6}[2+3ln(\frac{\Lambda^2}{m_Q^2})]
-\frac{3m_Q^3}{4M^4}[-3+\gamma_E+3ln(\frac{\Lambda^2}{m_Q^2})]\Bigg)+\frac{1}{4M^2}[m_Q^2(1+6\beta+\beta^2)
-12m_Qm_s(3-4\beta+\beta^2)\nonumber\\&&-4u_0^2m_\pi^2(1+7\beta+\beta^2)]\Bigg\}A(u_0)+m_su_0\mu_\pi
(\tilde{\mu}_\pi^2-1)(-5+2\beta+3\beta^2)\Bigg\{-\frac{10M^2\gamma_E}{m_Q^2}+1+2\gamma_E+ln(\frac{\Lambda^2}{m_Q^2})
\nonumber\\&&+\frac{m_Q^2}{6M^2}[2+3ln(\frac{\Lambda^2}{m_Q^2})]\Bigg\}\varphi_{\sigma}'(u_{0})+\frac{f_\pi
m_\pi^2
u_0}{2}(1+7\beta+\beta^2)A'(u_0)+\frac{6m_sM^2\gamma_E}{m_Q^3}\Bigg\{f_\pi
m_\pi^2(-1-4\beta+5\beta^2)(\eta_1-2\eta_2)\nonumber\\&&+6m_Q\mu_\pi(-5+2\beta+3\beta^2)(2\eta'_6-\eta'_5)\Bigg\}
+\frac{m_Q^2m_sm_\pi^2}{M^4}(\beta-1)\Bigg\{m_Qf_\pi\Bigg(u_0(3-\beta)[2+3ln(\frac{\Lambda^2}{m_Q^2})]\zeta\nonumber\\&&
+2[-5-\beta+(\beta-7)ln(\frac{\Lambda^2}{m_Q^2})](\eta_1-2\eta_2)\Bigg)+2\mu_\pi
u_0(5+3\beta)[2+3ln(\frac{\Lambda^2}{m_Q^2})](2\eta_{10}-\eta_9)\Bigg\}\nonumber\\&&
+\frac{1}{M^2}\Bigg\{m_Qf_\pi
m_\pi^2\Bigg(-m_Q(7+10\beta+7\beta^2)\eta_1-m_s[8(4-5\beta+\beta^2)
-3(-5+4\beta+\beta^2)\gamma_E(\eta_1-2\eta_2)\nonumber\\&&-7u_0(3-4\beta+\beta^2)\zeta]
+9m_s(3-4\beta+\beta^2)(\eta_1-2\eta_2)ln(\frac{\Lambda^2}{m_Q^2})\Bigg)
+2u_0^2f_\pi
m_\pi^4[3(1+4\beta+\beta^2)\eta_1+(1-14\beta+\beta^2)\eta_2]\nonumber\\&&
+m_Q^2m_s\mu_\pi(-5+2\beta+3\beta^2)[2+3ln(\frac{\Lambda^2}{m_Q^2})](\eta'_5-2\eta'_6)
+2u_0m_\pi^2\mu_\pi[m_Q(-1+4\beta-3\beta^2)\nonumber\\&&
-6m_s(-5+2\beta+3\beta^2)](2\eta_6-\eta_5)\Bigg\}+\frac{1}{m_Q^2}\Bigg\{m_Qf_\pi
m_\pi^2\Bigg(2m_Qu_0(1+7\beta+\beta^2)(\eta'_3-2\eta'_4)\nonumber\\&&
+m_Qu_0(1-14\beta+\beta^2)(\eta'_7-2\eta'_8)-m_Qu_0(3+22\beta+3\beta^2)\zeta
+m_s[12(1-2\beta+\beta^2)(\eta_1-\eta_2)+u_0(3-4\beta+\beta^2)\zeta]\Bigg)\nonumber\\&&
+[m_Q^3\mu_\pi(-1+4\beta-3\beta^2)-6m_Q^2m_s\mu_\pi(-5+2\beta+3\beta^2)](2\eta'_6-\eta'_5)
+2u_0m_\pi^2\mu_\pi[m_Q(2-8\beta+6\beta^2)\nonumber\\&&-3m_s(-5+2\beta+3\beta^2)]\Bigg\}\Bigg].
\end{eqnarray}
The other functions entering Eqs. (\ref{rho1}-\ref{gamma1}) are
given as
\begin{eqnarray}\label{etalar}
\eta_{i} &=& \int {\cal D}\alpha_i \int_0^1 dv f_{j}(\alpha_i)
\delta(\alpha_{ q} + v \alpha_g -  u_0),
\nonumber \\
\eta'_{i} &=& \int {\cal D}\alpha_i \int_0^1 dv f_{j}(\alpha_i)
\delta'(\alpha_{ q} + v \alpha_g -  u_0),
\nonumber \\
\psi_{nm}&=&\frac{{( {s-m_{Q}}^2 )
}^n}{s^m{(m_{Q}^{2})}^{n-m}},\nonumber \\
\end{eqnarray}
 and  $f_{1}(\alpha_i)={\cal V_{\parallel}}(\alpha_i)$, $f_{2}(\alpha_i)={\cal V_{\perp}}(\alpha_i)$,
 $f_{3}(\alpha_i)={\cal A_{\parallel}}(\alpha_i)$, $f_{4}(\alpha_i)=v{\cal A_{\parallel}}(\alpha_i)$, $f_{5}(\alpha_i)={\cal
 T}(\alpha_i)$, $f_{6}(\alpha_i)=v{\cal T}(\alpha_i)$, $f_{7}(\alpha_i)={\cal A_{\perp}}(\alpha_i)$
 and $f_{8}(\alpha_i)=v{\cal A_{\perp}}(\alpha_i)$ are the pion distribution amplitudes. Note that,
 in the above equations, the Borel parameter $M^2$  is defined as $M^{2}=\frac{M_{1}^{2}M_{2}^{2}}{M_{1}^{2}+M_{2}^{2}}$ and
$u_{0}=\frac{M_{1}^{2}}{M_{1}^{2}+M_{2}^{2}}$.  Since the mass of
the initial and final baryons are the same,
 we can set $ M_{1}^{2} = M_{2}^{2} $ and $u_{0} =\frac{1}{2}$.

 \section*{Appendix B}

\begin{eqnarray}\label{residurho1}
\rho_{1}(s)&=&\frac{1}{4096
\pi^4}\Bigg[4\Big\{m_Q^3m_s(1-4\beta+3\beta^2)+m_Q^3m_u(-7+4\beta+9\beta^2)\Big\}\Big\{6\psi_{10}
-3\psi_{20}+3\psi_{31}-\psi_{32}-6ln(\frac{s}{m_Q^2})\Big\}\nonumber\\&&
-(13+2\beta+13\beta^2)\Big\{12\psi_{10}
-6\psi_{20}+2\psi_{30}-4\psi_{41}+\psi_{42}-12\psi_{00}ln(\frac{s}{m_Q^2})\Big\}\Bigg]
+\frac{\langle\bar{u}u\rangle}{4096\pi^2}\Bigg[-\frac{8m_0^2}{m_Q}\Big\{10(\beta^2-1)(\psi_{00}\nonumber\\&&-\psi_{11})
-(-21+4\beta+17\beta^2)\psi_{02}\Big\}-32m_Q(-1+4\beta+3\beta^2)(2\psi_{10}-\psi_{11}-\psi_{12}+2\psi_{21})
\nonumber\\&&-\frac{8m_0^2}{m_Q^2}\Big\{16m_s(\beta^2-1)[2\gamma_E\psi_{00}-\psi_{01}-2\psi_{12}]
-m_s(-13+2\beta+11\beta^2)\psi_{02}-m_u(7+10\beta+7\beta^2)\psi_{02}\Big\}\nonumber\\&&
+4m_Q^3[m_s(1-4\beta+3\beta^2)+m_u(-7+4\beta+3\beta^2)]\Big\{6\psi_{10}-3\psi_{20}+3\psi_{31}-\psi_{32}-6ln(\frac{s}{m_Q^2})\Big\}\nonumber\\&&
-(13+2\beta+13\beta^2)m_Q^4\Big\{12\psi_{10}-6\psi_{20}+2\psi_{30}-4\psi_{41}+\psi_{42}+12ln(\frac{s}{m_Q^2})\psi_{00}\Big\}
\nonumber\\&&-\frac{16}{M^2}\Big\{M^2[2m_s(-7-2\beta+9\beta^2)-m_u(13+2\beta+13\beta^2)](\psi_{11}+\psi_{12})
-16m_0^2m_s(\beta^2-1)ln(\frac{s}{m_Q^2})\psi_{00}\Big\}\Bigg]\nonumber\\&&
+\frac{\langle\bar{s}s\rangle}{512}\Bigg[-\frac{m_0^2}{m_Q}\Big\{2(\beta^2-1)(\psi_{00}-\psi_{11})-(-5-4\beta+9\beta^2)\psi_{02}\Big\}
-4m_Q(1-4\beta+3\beta^2)\Big\{2\psi_{10}-\psi_{11}\nonumber\\&&-\psi_{12}+2\psi_{21}\Big\}
-\frac{m_0^2}{m_Q^2}\Big\{16m_u(\beta^2-1)(2\gamma_E\psi_{00}-\psi_{01}-2\psi_{12})+[m_s(1+6\beta+\beta^2)\nonumber\\&&
-m_u(-21+2\beta+19\beta^2)]\psi_{02}\Big\}+\frac{2}{M^2}\Big\{M^2[m_s(13+2\beta+13\beta^2)+2m_u(7+2\beta-9\beta^2)]\nonumber\\&&
-16m_0^2m_u(\beta^2-1)ln(\frac{s-m_Q^2}{\Lambda^2})\Big\}\Bigg]
+\frac{\langle
g^2G^2\rangle}{12288\pi^4}\Bigg[(19-34\beta+19\beta^2)(\psi_{11}+\psi_{12})+(46+44\beta+46\beta^2)\psi_{21}\nonumber\\&&
-9\Big\{(1-6\beta+\beta^2)(\psi_{11}+\psi_{21})+4(3+2\beta+3\beta^2)\psi_{21}+2(5+10\beta+5\beta^2)\psi_{10}\Big\}\nonumber\\&&
+\frac{2(\beta-1)}{m_Q}\Big\{m_s(23+3\beta)\psi_{02}+m_s(3\beta-1)(\psi_{10}-5\psi_{11}-2\psi_{12}+3\psi_{21})
+m_u(55+75\beta)\psi_{02}\nonumber\\&&+3m_u(7+3\beta)\psi_{10}+m_u(13+33\beta)\psi_{11}-m_u(14+6\beta)\psi_{12}+m_u(21+9\beta)\psi_{21}
-6(1+\beta)(3m_s+7m_u)\psi_{01}\nonumber\\&&+3[m_s(-1+3\beta+4(1+\beta)\gamma_E)-m_u(9+13\beta-4(1+\beta)\gamma_E)]\psi_{00}
+[m_s(\psi_{01}-3\psi_{02})\nonumber\\&&+m_u(5\psi_{01}+7\psi_{02})][ln(\frac{s^2-sm_Q^2}{\Lambda^2m_Q^2})
+ln(\frac{s-m_Q^2}{\Lambda^2})-ln(\frac{s}{\Lambda^2})]\Big\}\Bigg]\nonumber\\&&
+\frac{\langle
g^2G^2\rangle(m_s\langle\bar{u}u\rangle+m_u\langle\bar{s}s\rangle)}{3072\pi^2}
\Bigg[(7+2\beta-9\beta^2)\Big\{\frac{(m_0^2m_Q^4-4m_Q^2M^4)}{M^{10}}ln(\frac{s-m_Q^2}{\Lambda^2})\nonumber\\&&
+\frac{2}{m_Q^4}[(2\psi_{12}+\psi_{13}+11\psi_{14})-6ln(\frac{s-m_Q^2}{\Lambda^2})\psi_{00}
+6ln(\frac{s^2-sm_Q^2}{\Lambda^2m_Q^2})]\Big\}\Bigg]\nonumber\\&&+\frac{\langle
g^2G^2\rangle \langle\bar{u}u\rangle}{1536\pi^2}
\Bigg[m_s\Big\{(5-18\beta+13\beta^2)\frac{m_0^2m_Q^2}{12M^8}-(31-2-33\beta^2)\frac{1}{M^4}
\nonumber\\&&-(19+2\beta-21\beta^2)\frac{1}{m_Q^2M^2}\Big\}ln(\frac{s-m_Q^2}{\Lambda^2})\psi_{00}
-6(\beta^2-1)\frac{(m_Q-2m_s)}{m_Q^4}\psi_{02}\Bigg]\nonumber\\&&
+\frac{\langle g^2G^2\rangle
\langle\bar{s}s\rangle}{1536\pi^2}\Bigg[m_u\Big\{-(19+18\beta-37\beta^2)\frac{m_0^2m_Q^2}{12M^8}-(15+2-17\beta^2)\frac{1}{M^4}
\nonumber\\&&-(11-24\beta+13\beta^2)\frac{1}{m_Q^2M^2}\Big\}ln(\frac{s-m_Q^2}{\Lambda^2})\psi_{00}
+2(\beta^2-1)\frac{(m_Q-2m_u)}{m_Q^4}\psi_{02}\Bigg],\nonumber\\
\end{eqnarray}

\begin{eqnarray}\label{lamgamma1}
\Gamma_{1}&=&\frac{m_0^2(m_s\langle\bar{u}u\rangle+m_u\langle\bar{s}s\rangle)}{16\pi^2}\Bigg[\frac{
(\beta^2-1)\gamma_E
M^2}{m_Q^2}-\frac{1}{96}\Big\{6(7+2\beta-9\beta^2)+96(\beta^2-1)\gamma_E
-48(\beta^2-1)ln(\frac{m_Q^2}{\Lambda^2})\Big\}\Bigg]\nonumber\\&&
-\frac{5m_0^2(m_u\langle\bar{u}u\rangle+m_s\langle\bar{s}s\rangle)}{1536\pi^2}(1+\beta^2)
+\frac{\langle\bar{u}u\rangle
\langle\bar{s}s\rangle}{1152M^2}\Bigg[\frac{5m_0^2m_Q^3}{M^4}\Big\{m_s(-7+4\beta+3\beta^2)
+m_u(1-4\beta+3\beta^2)\Big\}\Bigg]\nonumber\\&&+\frac{12m_0^2m_Q^2}{M^2}(7+2\beta-9\beta^2)
+6m_0^2(15+2\beta-17\beta^2)-24M^2(7+2\beta-9\beta^2)-\frac{6m_Q(\beta-1)}{M^2}\Big\{m_sm_0^2(3+\beta)\nonumber\\&&
+m_um_0^2(1+3\beta)+2m_sM^2(7+3\beta)+2m_uM^2(3\beta-1)\Big\}\Bigg]-\frac{\langle
g^2G^2\rangle}{512m_Q\pi^4}(\beta^2-1)(m_u+m_s)M^2\gamma_E\nonumber\\&&
+\frac{\langle\bar{u}u\rangle \langle
g^2G^2\rangle}{18432\pi^2}\Bigg[\frac{m_0^2m_Q^4m_s}{M^8}\Big\{2-3ln(\frac{m_Q^2}{\Lambda^2})\Big\}
+\frac{m_0^2m_Q^2}{6M^6}\Big\{6m_s(67+18\beta-85\beta^2)+5m_u(1+2\beta+\beta^2)\nonumber\\&&
-3m_s(5-18\beta+13\beta^2)ln(\frac{m_Q^2}{\Lambda^2})\Big\}-\frac{4}{m_Q^2}\Big\{4m_Q(1+2\beta-3\beta^2)
+6m_s[7+2\beta-9\beta^2+(19+2\beta-21\beta^2)\gamma_E]\nonumber\\&&-3m_s(19+2\beta-21\beta^2)ln(\frac{m_Q^2}{\Lambda^2})\Big\}
+\frac{1}{2M^4}\Big\{16m_Q^2m_s(-7-2\beta+9\beta^2)+5m_0^2m_Q(1+4\beta-5\beta^2)\nonumber\\&&-136m_0^2m_s(\beta^2-1)
+24m_Q^2m_s(7+2\beta-9\beta^2)ln(\frac{m_Q^2}{\Lambda^2})\Big\}+\frac{1}{2m_Q^2M^2}\Big\{2m_0^2m_Q(-11+4\beta+7\beta^2)\nonumber\\&&
+m_0^2m_s(89+6\beta-95\beta^2)-24m_Q^2m_s(7+2\beta-9\beta^2)(2+\gamma_E)-76m_Q^2m_u(1+\beta)^2+12m_Q^2m_u(1+6\beta+\beta^2)\nonumber\\&&
+24m_Q^2m_s(19+2\beta-21\beta^2)ln(\frac{m_Q^2}{\Lambda^2})\Big\}\Bigg]
+\frac{\langle\bar{s}s\rangle \langle
g^2G^2\rangle}{18432\pi^2}\Bigg[\frac{m_0^2m_Q^4m_u}{M^8}\Big\{2-3ln(\frac{m_Q^2}{\Lambda^2})\Big\}\nonumber\\&&
+\frac{m_0^2m_Q^2}{6M^6}\Big\{18m_u(-25-6\beta+31\beta^2)+5m_s(37+50\beta+37\beta^2)
-3m_u(19+18\beta-37\beta^2)ln(\frac{m_Q^2}{\Lambda^2})\Big\}\nonumber\\&&+\frac{4}{m_Q^2}\Big\{4m_Q(1+2\beta-3\beta^2)
-6m_u[7+2\beta-9\beta^2+(11+2\beta-13\beta^2)\gamma_E]+3m_u(11+2\beta-13\beta^2)ln(\frac{m_Q^2}{\Lambda^2})\Big\}\nonumber\\&&
+\frac{1}{2M^4}\Big\{16m_Q^2m_u(-7-2\beta+9\beta^2)-m_0^2m_Q(3+20\beta-23\beta^2)-88m_0^2m_u(\beta^2-1)\nonumber\\&&
+24m_Q^2m_u(7+2\beta-9\beta^2)ln(\frac{m_Q^2}{\Lambda^2})\Big\}+\frac{1}{2m_Q^2M^2}\Big\{2m_0^2m_Q(-3-4\beta+7\beta^2)
+m_0^2m_u(65+6\beta-71\beta^2)\nonumber\\&&-24m_Q^2m_u(7+2\beta-9\beta^2)(2+\gamma_E)
+116m_Q^2m_s(1+\beta)^2-12m_Q^2m_s(7+10\beta+7\beta^2)\nonumber\\&&
+24m_Q^2m_u(11+2\beta-13\beta^2)ln(\frac{m_Q^2}{\Lambda^2})\Big\}\Bigg]
+\frac{\langle\bar{u}u\rangle \langle\bar{s}s\rangle \langle
g^2G^2\rangle}{13824M^6}\Bigg[\frac{5m_0^2m_Q^5}{M^6}\Big\{m_s(7-4\beta-3\beta^2)\nonumber\\&&
-m_u(1-4\beta+3\beta^2)\Big\}+\frac{m_0^2m_Q^3}{M^4}\Big\{2m_Q(-7-2\beta+9\beta^2)+5m_u(1+2\beta-3\beta^2)
-5m_s(7-4\beta-3\beta^2)\Big\}\nonumber\\&&+\frac{m_Q}{M^2}\Big\{m_Q^2m_s(-7+4\beta+3\beta^2)+m_Q^2m_u(1-4\beta+3\beta^2)
+4m_0^2m_Q(7+2\beta-9\beta^2)-5m_0^2m_u(1+2\beta-3\beta^2)\nonumber\\&&-5m_s(7-4\beta-3\beta^2)\Big\}-4m_Q^2(7+2\beta-9\beta^2)
+3m_Qm_u(1-4\beta+3\beta^2)-3m_Qm_s(7-4\beta-3\beta^2)\Bigg].
\end{eqnarray}

\clearpage
\begin{figure}[h!]
\begin{center}
\includegraphics[width=12cm]{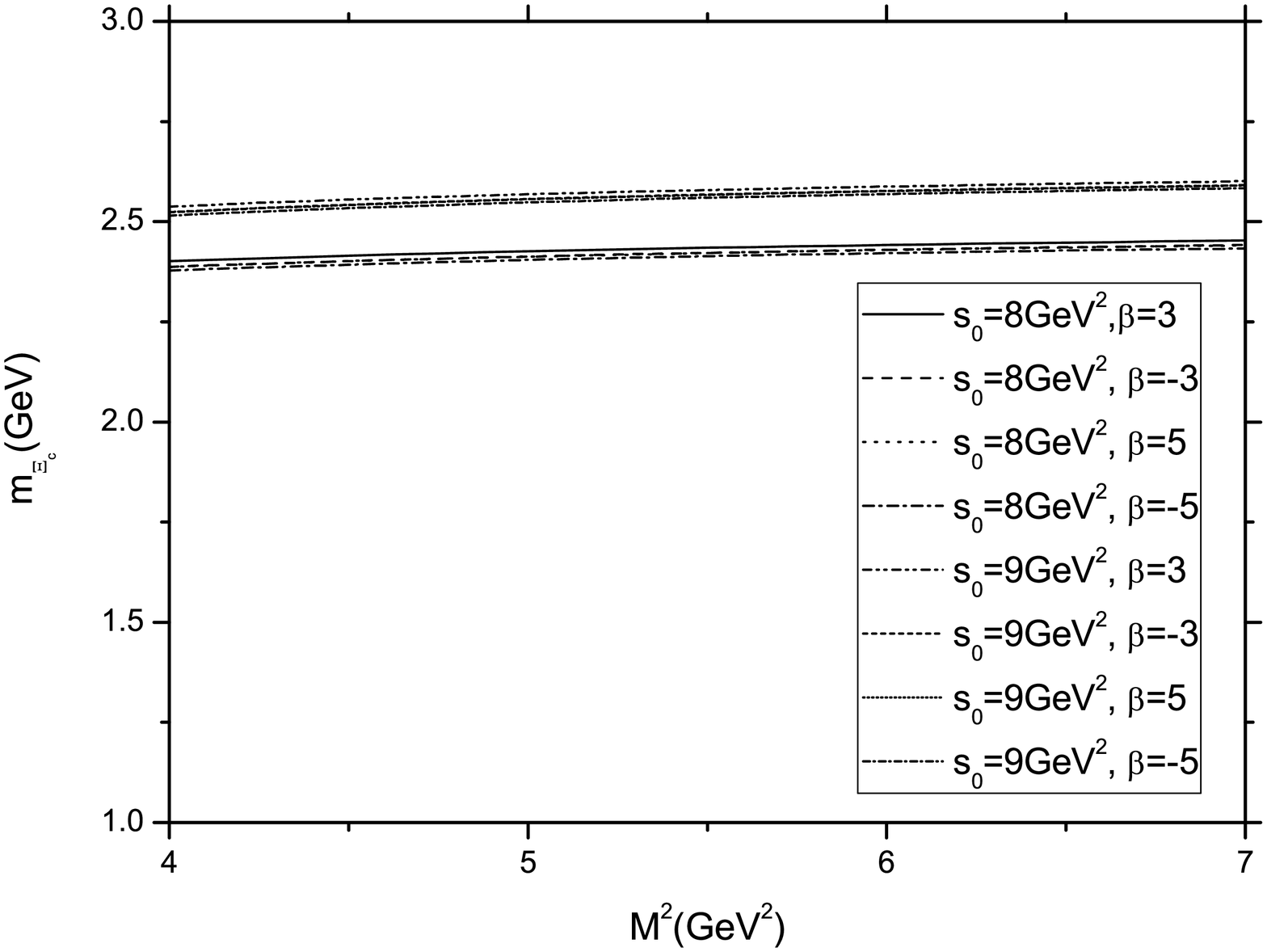}
\end{center}
\caption{The mass of the $\Xi_{c}$ as a function of the Borel
parameter $M^{2}$ for different values of arbitrary parameter
$\beta$ and the continuum threshold $s_{0}$.} \label{fig1}
\end{figure}

\begin{figure}[h!]
\begin{center}
\includegraphics[width=12cm]{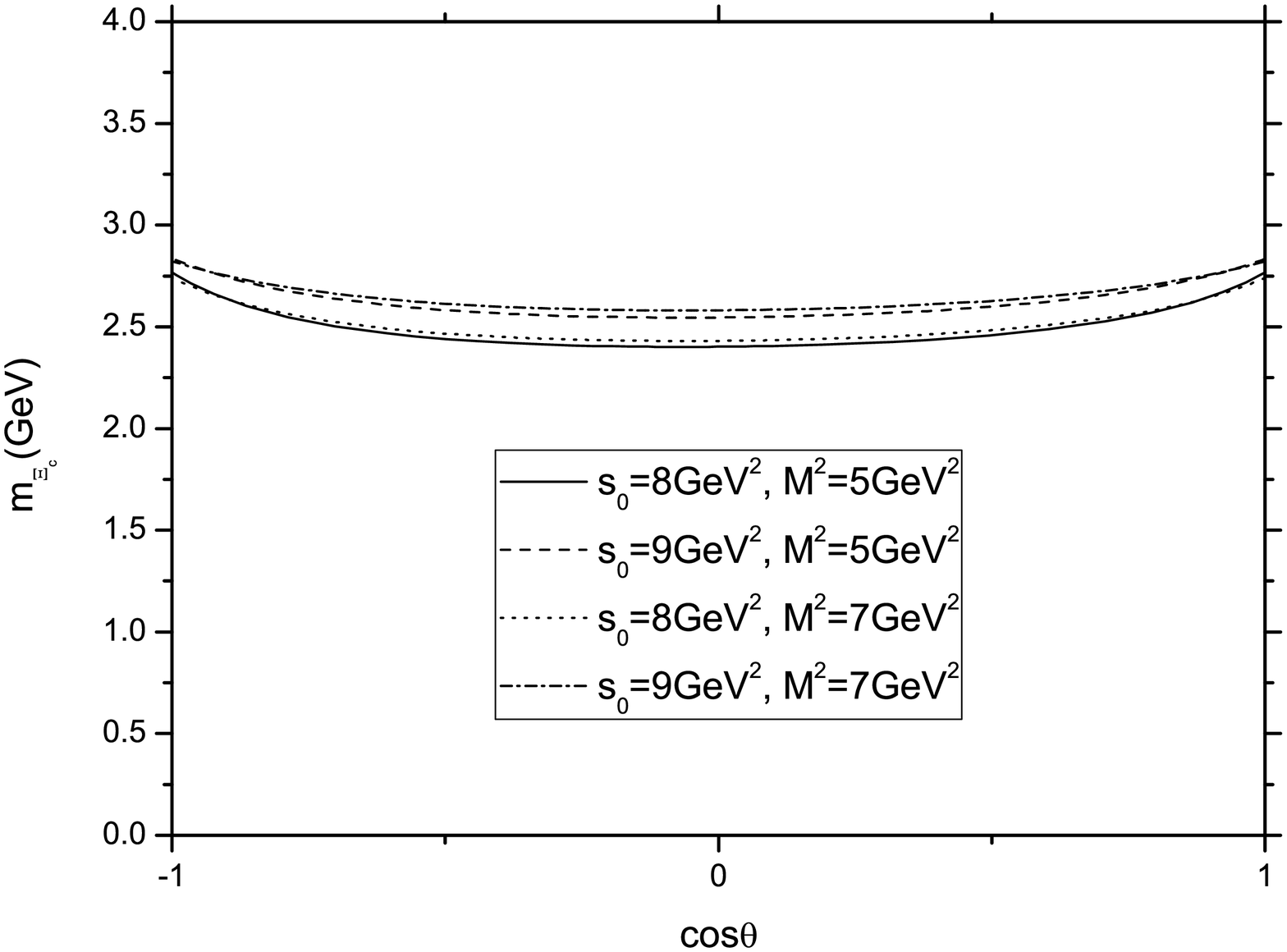}
\end{center}
\caption{The dependence of the mass of the $\Xi_{c}$ on $\cos\theta$
for different values of Borel parameter $M^{2}$ and the continuum
threshold $s_{0}$.} \label{fig2}
\end{figure}

\begin{figure}[h!]
\begin{center}
\includegraphics[width=12cm]{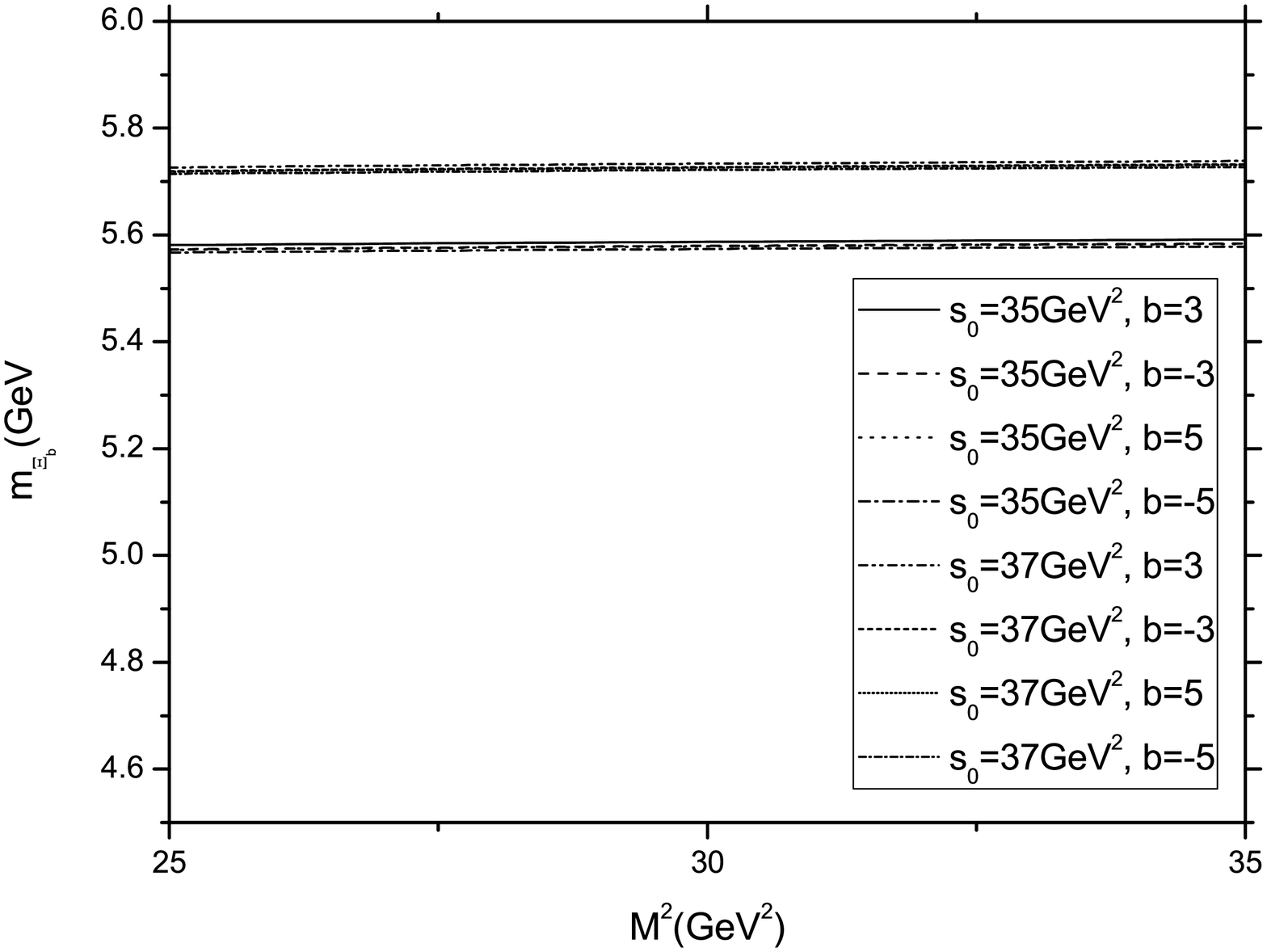}
\end{center}
\caption{The mass of the $\Xi_{b}$ as a function of the Borel
parameter $M^{2}$ for different values of arbitrary parameter
$\beta$ and the continuum threshold $s_{0}$.} \label{fig3}
\end{figure}

\begin{figure}[h!]
\begin{center}
\includegraphics[width=12cm]{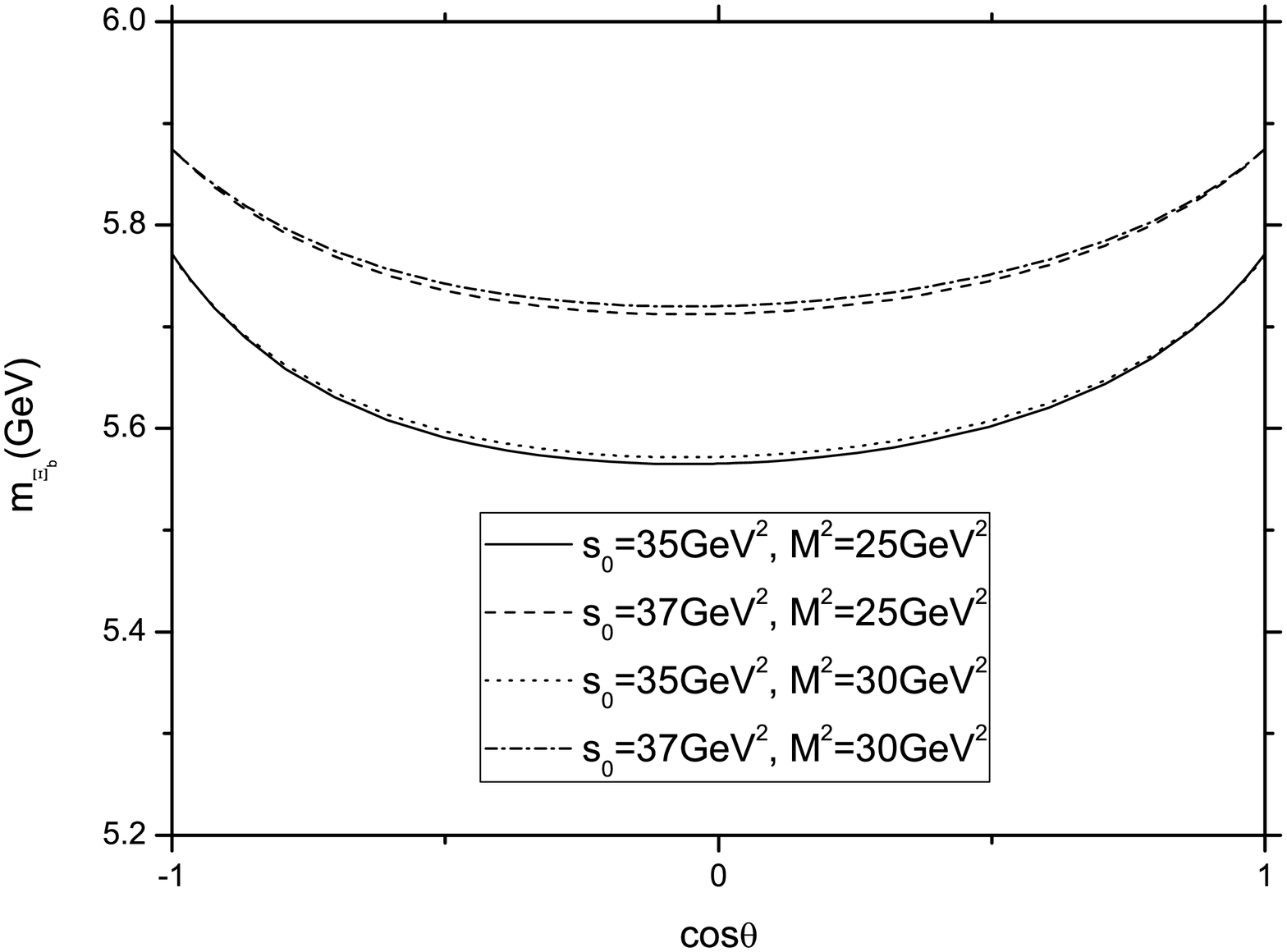}
\end{center}
\caption{The dependence of the mass of the $\Xi_{b}$ on $\cos\theta$
for different values of Borel parameter $M^{2}$ and the continuum
threshold $s_{0}$.} \label{fig4}
\end{figure}

\begin{figure}[h!]
\begin{center}
\includegraphics[width=12cm]{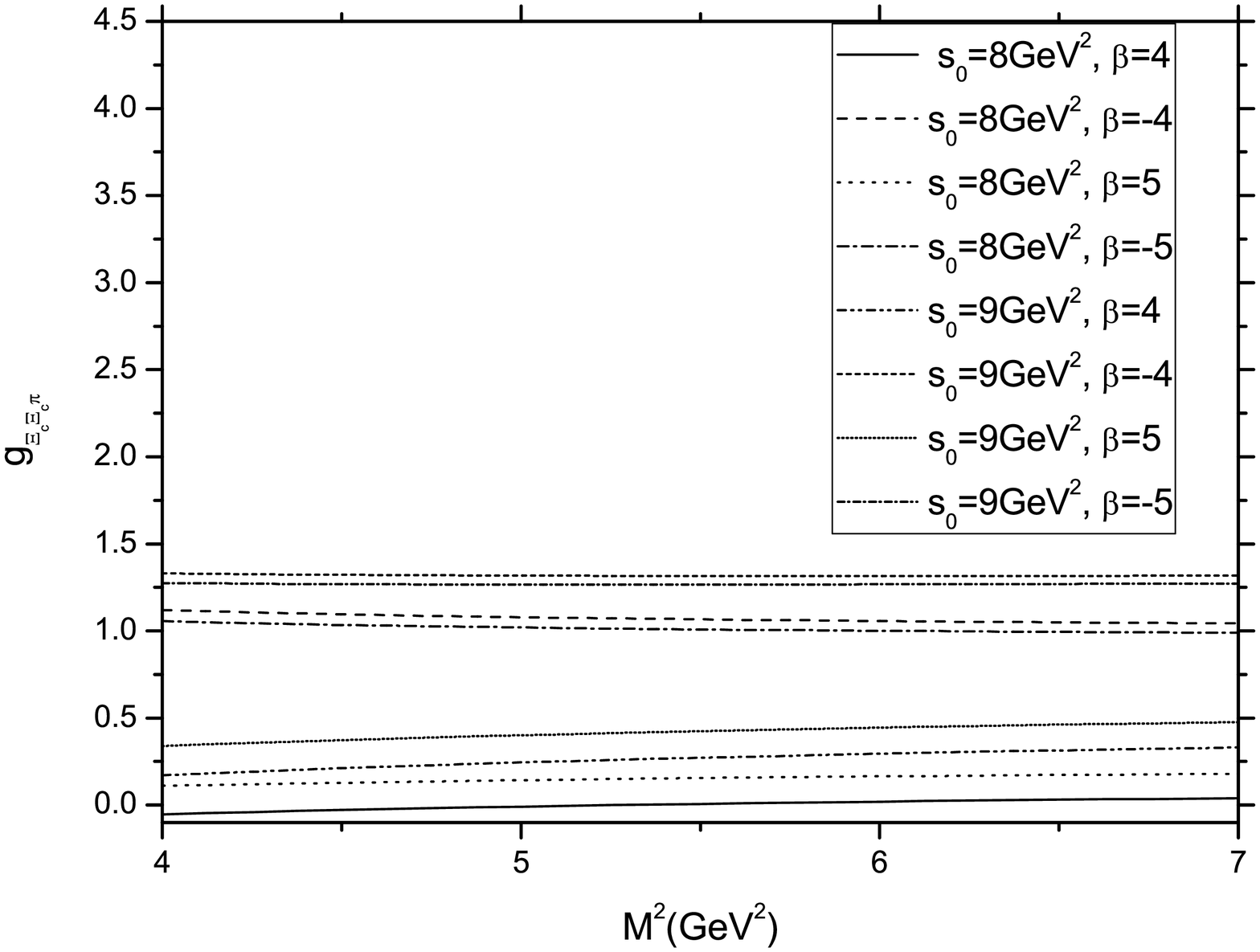}
\end{center}
\caption{The dependence of the coupling constant
$g_{{\Xi_{c}\Xi_{c}\pi}}$ on the Borel parameter $M^{2}$ for
different values arbitrary parameter $\beta$ and the continuum
threshold $s_{0}$.} \label{fig5}
\end{figure}

\begin{figure}[h!]
\begin{center}
\includegraphics[width=12cm]{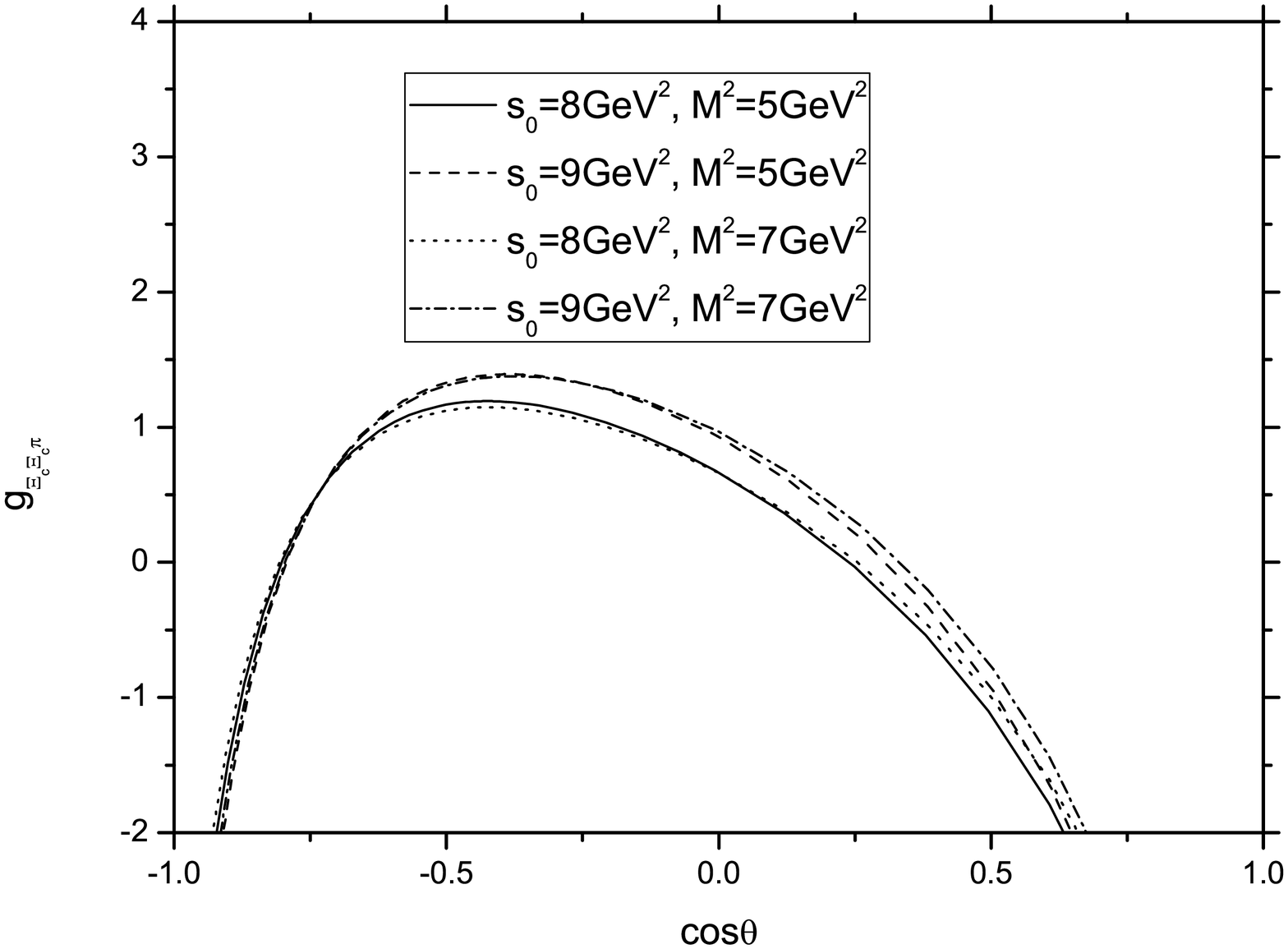}
\end{center}
\caption{The dependence of the coupling constant
$g_{{\Xi_{c}\Xi_{c}\pi}}$ on $\cos\theta$ for different values of
Borel parameter $M^{2}$ and the continuum threshold $s_{0}$.}
\label{fig6}
\end{figure}

\begin{figure}[h!]
\begin{center}
\includegraphics[width=12cm]{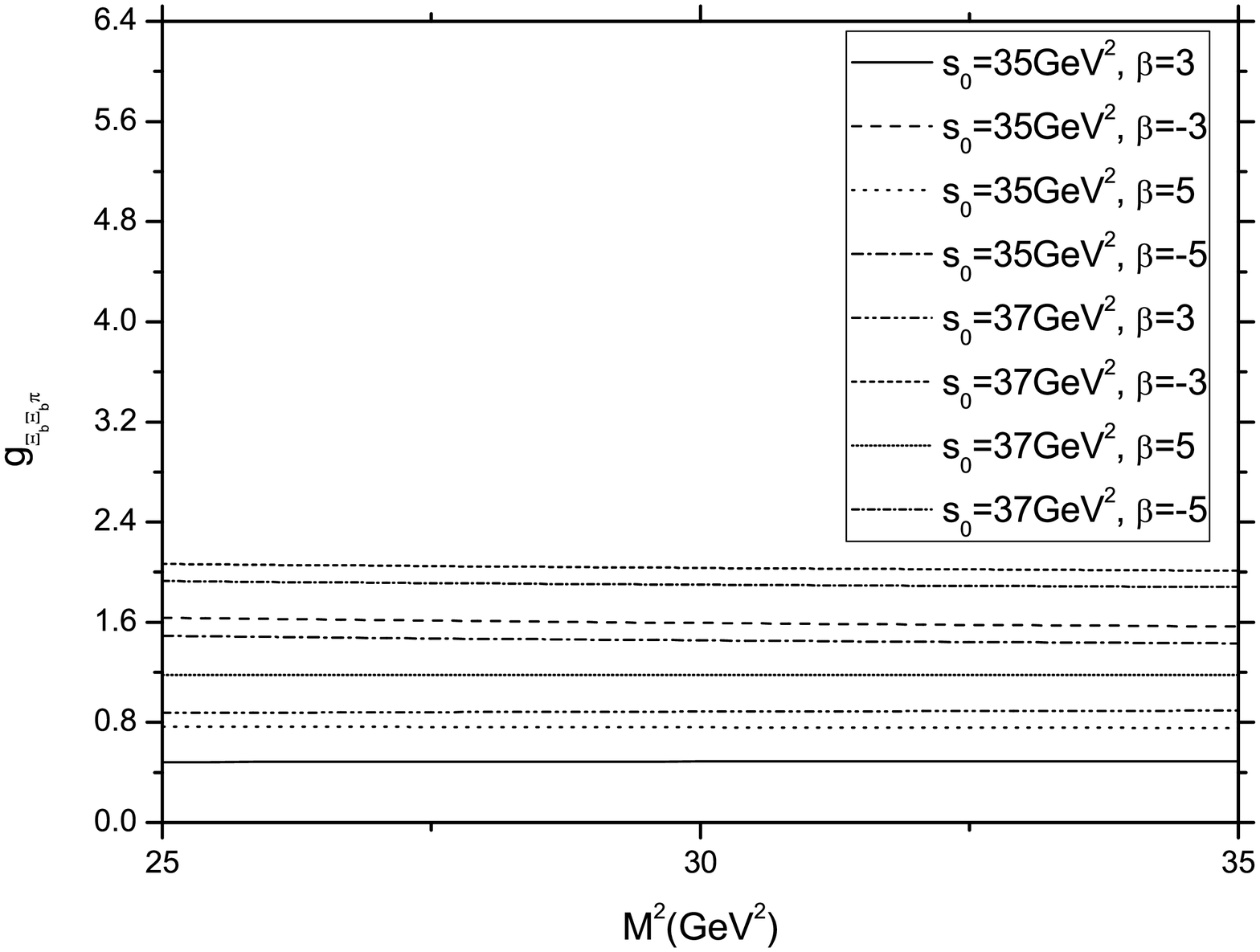}
\end{center}
\caption{The dependence of the coupling constant
$g_{{\Xi_{b}\Xi_{b}\pi}}$ on the Borel parameter $M^{2}$ for
different values of arbitrary parameter $\beta$ and the continuum
threshold $s_{0}$.} \label{fig7}
\end{figure}

\begin{figure}[h!]
\begin{center}
\includegraphics[width=12cm]{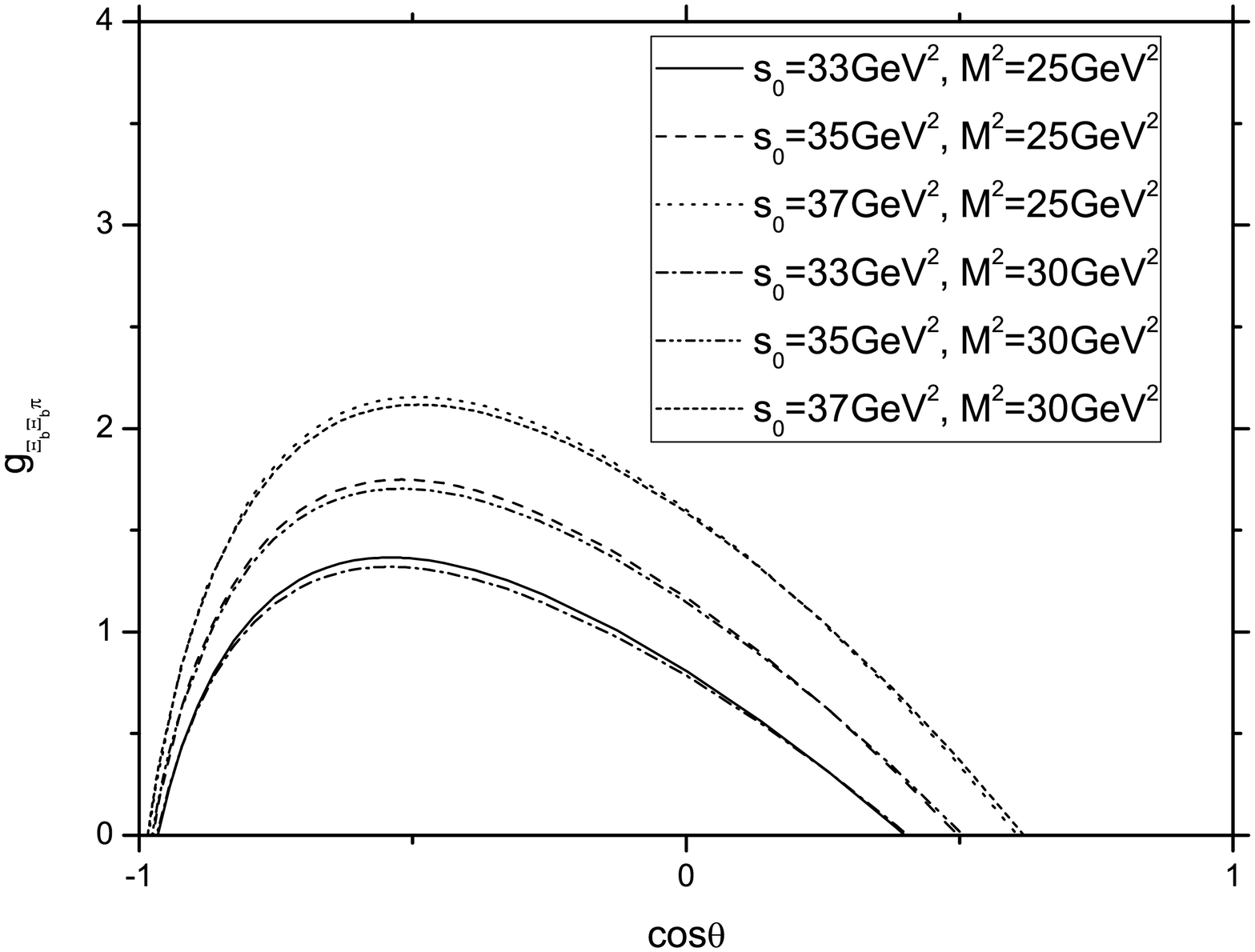}
\end{center}
\caption{The dependence of the coupling constant
$g_{{\Xi_{b}\Xi_{b}\pi}}$ on $\cos\theta$ for different values of
Borel parameter $M^{2}$ and the continuum threshold $s_{0}$.}
\label{fig8}
\end{figure}


\begin{thebibliography}{99}
\bibitem{Mizuk2005} R. Mizuk \emph{et. al}, Belle Collaboration, Phys. Rev. Lett. \textbf{94}, (2005)
122002.
\bibitem{Mizuk2007} R. Mizuk  \emph{et. al}, Belle Collaboration, Phys. Rev. Lett.
\textbf{98}, (2007) 262001  .
\bibitem{Aubert2007} B. Aubert  \emph{et. al}, BABAR Collaboration, Phys. Rev. Lett. \textbf{97}, (2006)
232001
\bibitem{Feindt2007} M. Feindt  \emph{et. al}, DELPHI Collaboration, Report no. CERN-PRE/95-139 (2007).
\bibitem{Edwards1995} K. W. Edwards  \emph{et. al}, CLEO Collaboration, Phys. Rev. Lett. \textbf{74}, (1995) 3331.
\bibitem{Artuso2001} M. Artuso  \emph{et. al}, CLEO Collaboration, Phys. Rev. Lett. \textbf{86}, (2001) 4479 .
\bibitem{I.V.Gorelov2007}  I. V. Gorelov CDF Collaboration, arXiv: 0701056 (hep-ex).
\bibitem{Aaltonen2007} T. Aaltonen \emph{et. al}, CDF Collaboration, Phys. Rev. Lett. \textbf{99}, (2007) 202001;
 T. Aaltonen \emph{et. al}, CDF Collaboration, Phys. Rev. Lett. \textbf{99}, (2007) 202002.
\bibitem{Abazov2007} V. Abazov \emph{et. al}, DO Collaboration, Phys. Rev. Lett. \textbf{99}, (2007) 052001.
\bibitem{Capstick}  S. Capstick, N. Isgur,  Phys. Rev. D
{\bf34} (1986) 2809.
\bibitem{Roncaglia}  R. Roncaglia, D. B. Lichtenberg, E. Predazzi,   Phys. Rev.
D {\bf52} (1995) 1722.
\bibitem{Jenkins}  E.  Jenkins, Phys. Rev.
D {\bf54} (1996) 4515.
\bibitem{Mathur}  N.  Mathur, R. Lewis , R. M.  Woloshyn,
Phys. Rev. D {\bf66} (2002) 014502.
\bibitem{Ebert}  D. Ebert, R. N. Faustov,  V. O.   Galkin, Phys. Rev.
D {\bf72} (2005) 034026.
\bibitem{Karliner1}  M. Karliner, H. J. Lipkin, arXiv:
0307343 (hep-ph), 0611306 (hep-ph).
\bibitem{Karliner2}  M. Karliner, B. Kereu-Zura, H. J. Lipkin, J. L. Rosner,  arXiv:
0706.2163 (hep-ph).
\bibitem{Rosner}  J. L.  Rosner, Phys. Rev.
D {\bf75} (2007) 013009.
\bibitem{Karliner3}  M. Karliner, H. J. Lipkin, Phys. Lett. B {\bf575} (2003) 249.
\bibitem{Shifman}  M. A. Shifman, A. I.  Vainshtein, V. I. Zakharov,
Nucl. Phys.  B {\bf 147} (1979) 385.
\bibitem{Bagan}  E. Bagan, M. Chabab, H. G. Dosch, S. Narison, Phys. Lett. B {\bf278} (1992) 367; ibid; Phys. Lett. B {\bf287} (1992) 176; ibid;
 Phys. Lett. B {\bf301} (1993) 243.
\bibitem{Navarra} C. S. Navarra, M. Nielsen, Phys. Lett. B {\bf 443} (1998) 285.
\bibitem{Shuryak}  E. V. Shuryak, Nucl. Phys. B
 {\bf 198} (1982) 83.
\bibitem{Grozin}  A. G.  Grozin, O. I. Yakovlev,  Phys.
Lett. B {\bf 285} (1992) 254.
\bibitem{Dai}  Y. B. Dai, C. S. Huang, C. Liu, C. D. Lu, Phys. Lett. B {\bf371} (1996)
99.
 \bibitem{Wang1} D. W.  Wang,  M. Q. Huang, C. Z. Li, Phys. Rev.  D  {\bf 65} (2002) 094036.
\bibitem{Zhu1} S. L. Zhu,  Phys. Rev. D {\bf 61} (2000) 114019.
\bibitem{Huang} C. S. Huang, A. L. Zhang, S. L. Zhu, Phys. Lett. B {\bf492} (2000)
288.
\bibitem{Wang2} D. W.  Wang,  M. Q. Huang,  Phys. Rev.  D  {\bf 68} (2003) 034019.
\bibitem{Wang3} Z. G.  Wang, Eur. Phys. J.  C  {\bf 54} (2008) 231.
\bibitem{Duraes} F. O.  Duraes, M. Nielsen, Phys. Lett. B  {\bf 658} (2007) 40.
\bibitem{Liu} X.  Liu, H. X. Chen, Y. R. Liu, A. Hosaka, S. L. Zhu, Phys. Rev.  D  {\bf 77} (2008) 014031.
\bibitem{Choudhury} A. L.  Choudhury,  V. Joshi,  Phys. Rev.  D  {\bf 13} (1976) 3115.
\bibitem{Lic} D. B.  Lichtenberg,   Phys. Rev.  D  {\bf 15} (1977) 345.
\bibitem{Glozman} L. Y.  Glozman,  D. O. Riska,  Nucl. Phys.  A  {\bf603} (1996) 326.
\bibitem{Julia} B.  Julia-Diaz,  D. O. Riska,  Nucl. Phys.  A  {\bf739} (2004) 69.
\bibitem{Schollamd} S.  Scholl,  H. Weigel,  Nucl. Phys.  A  {\bf735} (2004) 163.
\bibitem{Faessler} A.   Faessler et. al,  Phys. Rev.  D  {\bf 73} (2006) 094013.
\bibitem{Patel}  B. Patel, A. K. Rai, P. C. Vinodkumar, arXiv:
0803.0221 (hep-ph).
\bibitem{Savage} M. Savage, Phys. Lett. B  {\bf 326} (1994) 303.
\bibitem{DOR}  D. O. Riska, Nucl. Instrum. Meth. B {\bf 119} (1996) 259.
\bibitem{Oh} Y. Oh, D. P. Min, M. Rho, N. N. Scoccola, Nucl. Phys.  A  {\bf534} (1991) 493.
\bibitem{An} C. S. An, Nucl. Phys.  A  {\bf797} (2007) 131, Erratum-ibid; A  {\bf801} (2008) 82.
\bibitem{Zhu2} S. L. Zhu, W. Y. P. Hwang, Z. S. Yang,  Phys. Rev. D {\bf 56} (1997) 7273.
\bibitem{Aliev1} T. M.  Aliev, A. Ozpineci, M. Savci, Phys. Rev. D {\bf 65} (2002) 096004.
\bibitem{Aliev2} T. M.  Aliev, A. Ozpineci, M. Savci, Phys. Rev. D {\bf 65} (2002) 056008.
\bibitem{Azizi}  K. Azizi, M. Bayar, A. Ozpineci, Phys. Rev. D {\bf 79}
(2009) 056002.
\bibitem{Balitsky}
I. I. Balitsky, V. M. Braun, Nucl. Phys. B 311 (1989) 541.
\bibitem{R21}
P. Ball, JHEP {\bf 01} (1999) 010; P.Ball, V. Braun, A. Lenz, JHEP
{\bf 0605} (2006) 004.
\bibitem{R22} P. Ball, R. Zwicky, Phys. Rev. D {\bf 71}
(2005) 014015.
\bibitem{Belyaev} V. M. Belyaev,  B. L.  Ioffe, JETP  {\bf56} (1982) 493.








\end{thebibliography}
\end{document}